\documentclass[aps,pra,reprint,superscriptaddress,showpacs,longbibliography,floatfix,footnoteinbib]{revtex4-1}
\usepackage{amsmath,amssymb,graphicx,units}
\usepackage[plainpages=false,pdfpagelabels,colorlinks=true,linkcolor=red,
urlcolor=blue,citecolor=blue,pdftitle={Title},pdfauthor={},pdfdisplaydoctitle=true,
pdfduplex=DuplexFlipLongEdge]{hyperref}

\newcommand{\ket}[1]{{\vert #1\rangle}}

\newcommand{\braket}[2]{\langle#1\vert#2\rangle}

\newcommand{\1}{\mbox{\bf 1}}

\makeatother

\begin{document}
\title{Photoinduced anomalous Hall effect in the interacting Haldane model: targeting topological states with pump pulses}
\author{Can Shao}
\email{shaocan@njust.edu.cn}
\affiliation{Institute of Ultrafast Optical Physics, Department of Applied Physics, Nanjing University of Science and Technology, Nanjing 210094, China}

\author{P. D. Sacramento}
\affiliation{CeFEMA, Instituto Superior T\'{e}cnico, Universidade de Lisboa, Av. Rovisco Pais, 1049-001 Lisboa, Portugal}
\affiliation{Beijing Computational Science Research Center, Beijing 100193, China}

\author{Rubem Mondaini}
\email{rmondaini@csrc.ac.cn}
\affiliation{Beijing Computational Science Research Center, Beijing 100193, China}


\begin{abstract}
We investigate the nonequilibrium dynamics of the spinless Haldane model with nearest-neighbor interactions on the honeycomb lattice by employing an unbiased numerical method. In this system, a first-order transition from the Chern insulator (CI) at weak coupling to the charge-density-wave (CDW) phase at strong coupling can be characterized by a level crossing of the lowest energy levels. Here we show that adiabatically following the eigenstates across this level crossing, their Chern numbers are preserved, leading to the identification of a topologically-nontrivial low-energy excited state in the CDW regime. By promoting a resonant energy excitation via an ultrafast circularly polarized pump pulse, we find that the system acquires a non-vanishing Hall response as a result of the large overlap enhancement between the time-dependent wave-function and the topologically non-trivial excited state. This is suggestive of a photoinduced topological phase transition via unitary dynamics, despite a proper definition of the Chern number remaining elusive for an out-of-equilibrium interacting system. We contrast these results with more common quench protocols, where such features are largely absent in the dynamics even if the post-quench Hamiltonian displays a topologically nontrivial ground state.
\end{abstract}

\maketitle

\section{Introduction}\label{sec1}

Topological phases of matter have been largely understood and fully classified in terms of symmetries (time-reversal, particle-hole, chiral and crystal symmetries) of the quantum systems under investigation~\cite{Chiu16, Kruthoff17, Zhang19, Vergniory19, Tang19}. Such phases can be characterized by their topological invariants and are robust against weak time-reversal invariant perturbations~\cite{Bansil16}. In ultracold atomic systems, the preparation of topologically non-trivial states via dynamically engineered perturbations has been experimentally realized and measured~\cite{Jotzu14, Aidelsburger15}. However, it has been demonstrated that in the thermodynamic limit, topological characteristics of the time-dependent wave function are preserved under local unitary evolution~\cite{Chen10, Foster13, Sacramento14, Alessio15, Sacramento16, Caio16}. As a direct consequence, it puts into question how dynamical topological transitions can take place in isolated quantum systems.

To overcome such obstacle, approaches were proposed that include dephasing~\cite{Ying16} or the presence of interactions~\cite{Kruckenhauser18, Michael19}. The latter may allow a system to act as its own bath and stabilize a dynamically induced topological phase, despite still being governed by a unitary evolution. In particular, for two-dimensional systems in equilibrium, the celebrated TKNN expression~\cite{Thouless82} directly relates the mathematical topological invariant, the Chern number $C$, with a physical observable, the Hall response $\sigma_{xy}$. Departing from equilibrium conditions, growing evidence shows that such relation breaks down, that is, an induced Hall response may coexist with an invariant Chern number~\cite{Ying16,Wilson16,Michael19,Ge21,Peralta18}.

Our starting point is an interacting model in which the formation of a local order parameter directly competes with the topological phase~\cite{Varney10, Varney11}, part of a large scope of studies with similar scenarios~\cite{Rachel10, Yamaji11, Zheng11, Yu11, Griset12, Hohenadler11, Hohenadler12, Reuther12, Laubach14, Shao21}. These are to be contrasted with investigations wherein topological Mott insulators are argued to be directly induced by the interactions~\cite{Raghu08, Wen10, Weeks10, Ruegg11, Yang11, Budich12, Dauphin12, LeiWang12, Yoshida14}. The fundamental question we address is on the possibility that a non-trivial Hall response can be obtained when performing dynamical perturbations in the otherwise topologically trivial phase within the strong coupling regime. For that, we explore the route of promoting engineered perturbations, in particular on subjecting the system to circularly polarized light. This has been shown in the past, for non-interacting schemes, to induce a non-trivial topological response~\cite{Oka09, Inoue10, Kitagawa11, Lindner11, Perez-Piskunow14, Usaj14}.

Going beyond the usual Floquet picture of time-periodic drivings, our approach is connected to recent experiments that make use of a short-lived perturbation, i.e., a femtosecond pulse of circularly polarized light~\cite{McIver20}. Such method endows the ability of fine tuning the amount of energy deposited in the system and, consequently, resonantly explore some of its excited states~\cite{Shao19}. Unlike in experiments, since we do not possess explicit dephasing mechanisms, its effect on engineering non-trivial topological characteristics is long-lived, in spite of the perturbation being constrained in time.

In details, we study the half-filled spinless Haldane model with nearest-neighbor repulsion in- and out-of-equilibrium, by means of exact diagonalization (ED). The phase transition of this model, from a Chern insulator (CI) towards a trivial charge-density-wave (CDW) Mott insulator with growing interactions, has been studied by Varney \emph{et al.}~\cite{Varney10, Varney11}. They further demonstrated that clusters with sixfold rotational symmetry and $K$ point included in the discrete momentum space, display reduced finite-size effects. Hence, we adopt a 24-sites cluster [see inset in Fig.~\ref{fig_1}(b)] that is amenable to ED and satisfies the above conditions. Our study in equilibrium confirms that once a CDW order is formed after a level crossing (indicating the first-order phase transition), the topological characteristics of the ground state (GS) are no longer present.

We then notice that a low-energy excited state within the CDW regime, which is smoothly connected to the GS in the parent Chern insulating phase, preserves its finite Chern number and nonzero Hall response before merging into the continuum of the spectrum for even larger interactions. Based on the energy difference of this topological excited state to the GS, a circularly polarized pump is resonantly applied to stimulate the initial CDW ground state. We find that the overlap of the time-evolving wave function is oscillatory between the two states. The oscillating frequency mainly depends on the pump amplitude, resulting that the post-pump nonequilibrium state can be tuned by the laser strength. Finally, we contrast these results with the ones from a quench protocol, and we observe the inability to recover such dynamical topological phase.

The presentation is organized as follows: In Sec.~\ref{sec_model}, we introduce the model, methods and all the relevant quantities. An analysis of our results, in- and out-of-equilibrium, is shown in Sec.~\ref{sec_Results}, and a conclusion, accompanied by a discussion of the results, is given in Sec.~\ref{conclusion}.

\section{Model and method}\label{sec_model}

The model under investigation is the half-filled spinless Haldane model with repulsive nearest-neighbor interactions:
\begin{eqnarray}
\hat H=&-&t_1\sum_{\langle i,j\rangle}(\hat c^{\dagger}_{i} \hat c^{\phantom{}}_{j}+\text{H.c.})
-t_2\sum_{\langle\langle i,j\rangle\rangle}(e^{{\rm i}\phi_{ij}}\hat c^{\dagger}_{i} \hat c^{\phantom{}}_{j}+\text{H.c.}) \nonumber \\
&+&V\sum_{\langle i,j\rangle}\hat n_{i}\hat n_{j}.
\label{eq:H}
\end{eqnarray}
Here, $\hat c^{\dagger}_{i}$ ($\hat c^{\phantom{}}_{i}$) represents the fermionic creation (annihilation) operator at site $i$ and $\hat n_{i}$ is the corresponding number operator. $t_1$ ($t_2$) is the nearest-neighbor (next-nearest-neighbor) hopping constant and $V$ is the nearest-neighbor interaction. A phase $\phi_{i,j}=\frac{\pi}{2}$ ($-\frac{\pi}{2}$) in the anticlockwise (clockwise) loop is added to the second hopping term, which breaks the time-reversal symmetry, resulting in topologically non-trivial behavior in the system.

In equilibrium, we calculate the topological invariants of the GS and a selection of low-lying eigenstates. This is quantified by the Chern number, which is defined as an integration over the Brillouin zone~\cite{Niu85} using twisted boundary conditions~\cite{Didier91},
\begin{align}
  C = \int \frac{d\phi_x d\phi_y}{2 \pi i} \left( \langle\partial_{\phi_x}
      \Psi^\ast | \partial_{\phi_y} \Psi\rangle - \langle{\partial_{\phi_y}
      \Psi^\ast | \partial_{\phi_x} \Psi\rangle} \right),
\label{eq:C}
\end{align}
with $\ket{\Psi}$ the many-particle wave function, and $\phi_x$ ($\phi_y$) the twisted phase along the $x$ ($y$) direction. This continuous expression has been shown to converge to the correct one if using a sufficiently discretized version~\cite{Fukui05, Varney11}.

In out-of-equilibrium conditions, we employ the time-dependent Lanczos technique~\cite{Prelovsek} to evolve the many-body wave function via,
\begin{equation}
\ket{\psi(t+\delta{t})}=e^{-\mathrm{i}H(t)\delta t}\ket{\psi(t)}
\simeq\sum_{l=1}^{M}{e^{-\mathrm{i}\epsilon_l\delta{t}}}\ket{\phi_l}\braket{\phi_l}{\psi(t)},
\label{eq:lanczos}
\end{equation}
where $\epsilon_l$ and $|\phi_l\rangle$ are the eigenvalues and eigenvectors of the $M$-dimensional Krylov subspace generated in the Lanczos process at each instant of time $t$. Here, we chose $\delta{t}=0.02$ and $M=30$ to ensure the convergence of the numerical evolution. In what follows, the system can be excited by either a quench, with a sudden change of $V$, or by a pump pulse. In the latter, the external electric field during photoirradiation can be included into the Hamiltonian via the Peierls substitution in the hopping terms:
\begin{equation}
c^{\dagger}_{i,\sigma}c_{j,\sigma}+\text{H.c.}\rightarrow
e^{\mathrm{i}\textbf{A}(t)\cdot(\textbf{R}_j-\textbf{R}_i)}c^{\dagger}_{i,\sigma}c_{j,\sigma}+\text{H.c.},
\label{eq:Peierls}
\end{equation}
where,
\begin{equation}
\textbf{A}(t)=A_0e^{-\left(t-t_0\right)^2/2t_d^2} (\cos\left[\omega_0\left(t-t_0\right)\right],\sin\left[\omega_0\left(t-t_0\right)\right]),
\label{eq:vpotent}
\end{equation}
is the vector potential of a circularly polarized pump pulse. Its temporal envelope is centered at $t_0$ and taken to be Gaussian. The parameter $t_d$ controls its width, and $\omega_0$ is the central frequency. Additionally, we define $\Delta t=t-t_0$ as the time difference between the probing and pumping (or quench) instants.

To characterize the dynamics and the possible non-trivial Hall response, we calculate the following relevant quantities:\\
\indent i) $|\langle\psi(\Delta t)|\psi_n\rangle|$, the overlap between the time-dependent wave function and the $n$-th eigenstate of the equilibrium Hamiltonian [Eq.~\eqref{eq:H}] in the \textbf{k}=(0,0) quasi-momentum subspace. As the ground state is located at this momentum sector for our model parameters, and that both pump and quench scenarios do not break translational invariance, all the dynamics we explore is constrained to this subspace.\\
\indent ii) $S_{\text{CDW}} = \frac{1}{N}\sum\limits_{i,j}\langle (\hat n_{i}^{A}-\hat n_{i}^{B})(\hat n_{j}^{A}-\hat n_{j}^{B})\rangle$, structure factor of CDW, where $N=12$ is the number of unit cells and $\hat n_{i}^{\alpha}$ denotes the density number operator at site $i$ of sublattice $\alpha$ $= A$ or $B$.\\
\indent iii) $\tilde{\sigma}_{xy}(t_{\text{H}})$, the dynamical Hall response. By applying a weak electric field $E_x(t_{\text{H}})=F_0(1-e^{-t_{\text{H}}/\tau})$ along the $x$ direction, with $F_0=10^{-4}$ and $\tau=5$, we measure the induced current in the $y$ direction, $J_y(t_{\text{H}})$, as to obtain the dynamical Hall response $\tilde\sigma_{xy}(t_{\text{H}})=\frac{J_y(t_{\text{H}})}{F_0\cdot A_s}$, where $A_s$ is the total area of the cluster. In this process, $F_0$ is sufficiently small to mitigate its influence on the system's dynamics. Note that in what follows $E_x(t_{\text{H}})$ will be applied to either the eigenstates of the Hamiltonian~\eqref{eq:H}, such as the GS and second excited state, or to the time-dependent wave function $\psi(\Delta t)$ in out-of-equilibrium conditions. In the latter, we choose the reference probe time $t_{\text{H}}=0$ located at $\Delta t=40$ after the pump, when discussing the results of Fig.~\ref{fig_4}.
The electric field $E_x(t_{\text{H}})$ is introduced smoothly from $t_{\text{H}}=0$ by adding a vector potential
\begin{eqnarray}
A_{x}^{\rm Hall}(t_{\text{H}})=F_0(t_{\text{H}}+\tau\cdot e^{-t_{\text{H}}/\tau}-\tau).
\label{eq:Hall}
\end{eqnarray}
to Eq.~\eqref{eq:Peierls}. In summary, we employ three steps to obtain the Hall response after the pump:
\\ \indent 1. Apply a pump expressed by Eq.~\eqref{eq:vpotent}, calculating the current $J'_y(t_{\text{H}})$ from $\Delta t=40$;
\\ \indent 2. apply both the pump and weak electric field (starting from $\Delta t=40$) described by Eq.~\eqref{eq:vpotent} and Eq.~\eqref{eq:Hall}, respectively, to calculate $J^{''}_y(t_{\text{H}})$;
\\ \indent 3. obtain the net current $J_y(t_{\text{H}})=J^{''}_y(t_{\text{H}})-J^{'}_y(t_{\text{H}})$ so that $\tilde\sigma_{xy}(t_{\text{H}})=\frac{J_y(t_{\text{H}})}{F_0\cdot A_s}$ represents the nonequilibrium Hall response.

Alternatively, we can use the direct computation of the Kubo formula to benchmark $\tilde{\sigma}_{xy}(t_{\text{H}})$ in equilibrium:
\begin{equation}
\sigma_{xy}=\frac{{\rm i} \hbar}{A_s} \sum_{n \neq 0} \frac{\langle \psi_0|\hat J_y|\psi_n\rangle \langle \psi_n|\hat J_x|\psi_0\rangle-\langle \psi_0|\hat J_x|\psi_n\rangle \langle \psi_n|\hat J_y|\psi_0\rangle}{(E_n-E_0)^2}
\label{eq:lin_resp}
\end{equation}
where $(E_\alpha,\psi_\alpha)$ are the eigenpairs of the Hamiltonian~\eqref{eq:H}, and $\hat J_\mu$ is the current operator along the $\mu$ direction.

In what follows, we set the parameters $t_2=0.2$, using units where $e=\hbar=1$ and the lattice spacing $a_0=1$. In these units, $t_1$ and $t_1^{-1}$ are set to be the unit of energy and time, respectively.

\section{Results and Analysis}\label{sec_Results}

\begin{figure}[t]
\centering
\includegraphics[width=0.5\textwidth]{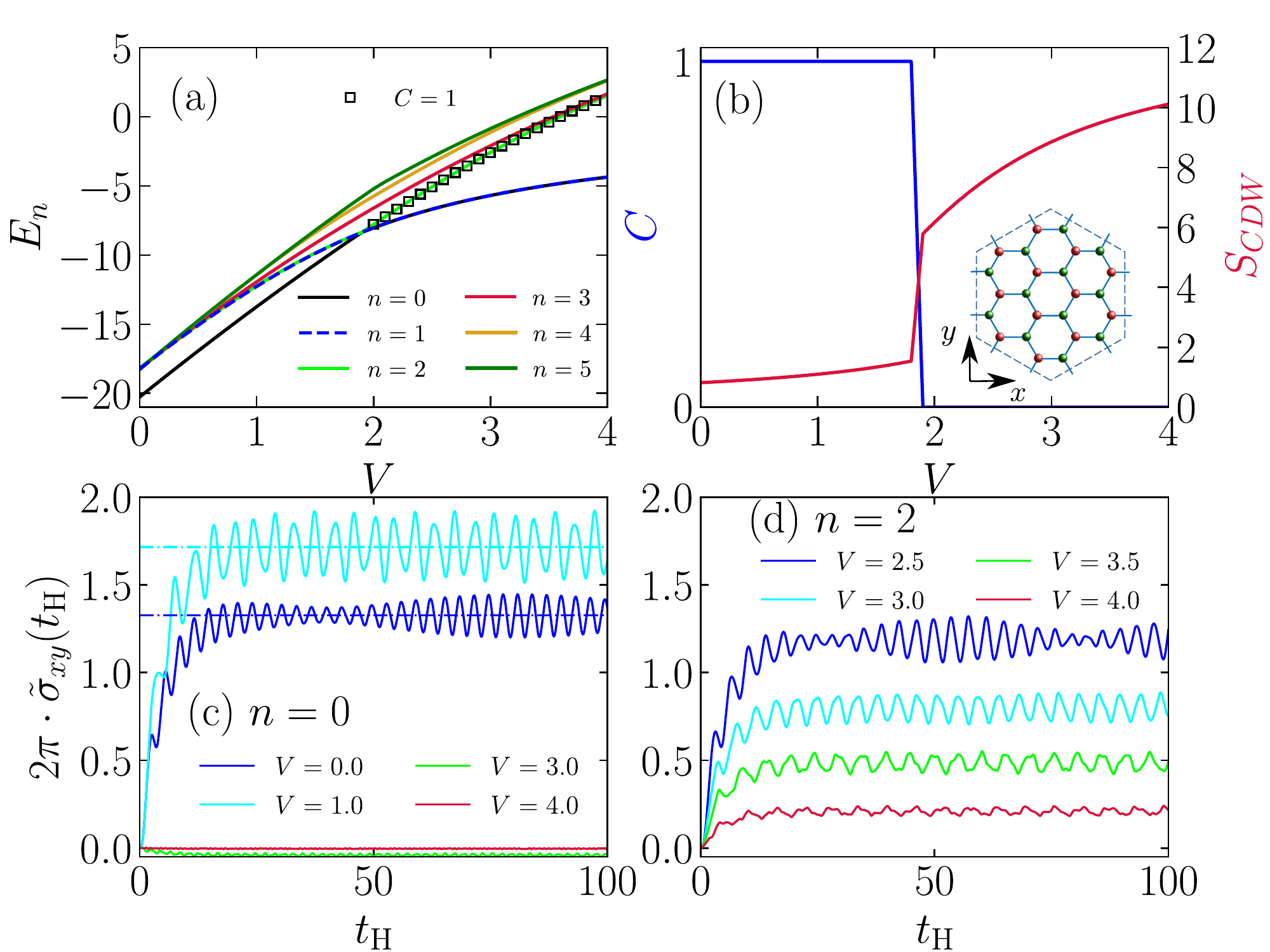}
\caption{Six lowest-lying energy levels (a), the ground-state Chern number and CDW structure factor (b) \emph{vs.} $V$ for the Hamiltonian \eqref{eq:H} in the $\textbf{k}=(0,0)$ subspace. The dynamical Hall response, $\tilde{\sigma}_{xy}(t_{\text{H}})$, of the GS with $V=0.0$, $1.0$, $3.0$, $4.0$ (c) and second excited state with $V=2.5$, $3.0$, $3.5$, $4.0$ (d), respectively. In (c), the equilibrium Hall response ${\sigma}_{xy}$ obtained from the Kubo formula, Eq.~\eqref{eq:lin_resp} (using a total of 2,000 low-lying eigenstates), is shown as horizontal dash-dotted lines. The cartoon in (b) depicts the 24-sites cluster.}
\label{fig_1}
\end{figure}

We start by calculating the low-lying energy spectrum versus the interaction strength $V$ for the Hamiltonian~\eqref{eq:H}, as shown in Fig.~\ref{fig_1}(a). A level crossing, associated with a first-order phase transition between the CI and CDW phases, locates at $V\simeq1.9$. The Chern number and CDW structure factor of the ground state also suddenly change at this point, see Fig.~\ref{fig_1}(b) -- these results are consistent with those in Ref.~[\onlinecite{Varney11}]. From Fig.~\ref{fig_1}(a), a close inspection reveals that the original CI ground state is smoothly connected to the \emph{second} excited state $|\psi_2\rangle$ when $V>1.9$ (green line), whereas the CDW ground state is nearly degenerate with the first excited state (dashed-blue line) in our finite lattice. We compute the Chern number of the $|\psi_2\rangle$ state and find that for $V\in[2,4]$ it always has $C=1$ [denoted by the black square markers in Fig.~\ref{fig_1}(a)]. The Chern numbers of other excited states can turn unstable: either not integers or different values for different $V$. The reason is attributed to the method to calculate Chern number, which requires that the manifold of eigenenergies of the state $|\psi_\alpha\rangle$ in the torus of $E_\alpha(\phi_x,\phi_y)$ to be always gapped~\cite{Fukui05,Varney11}, where $\phi_x$ and $\phi_y$ represent the twisted phases in two directions. Fortunately, the second excited state is always separated from the bulk of the spectrum in a finite range of interactions within the CDW phase, as one would expect from states that just underwent first order phase transitions~\cite{Chen10}.

To confirm the results of the Chern number, we check the dynamical Hall response, $\tilde{\sigma}_{xy}(t_{\text{H}})$, for the GS and the second excited state. For that, as mentioned in Sec.~\ref{sec_model}, a weak electric field along the $x$ direction is applied at time $t_{\text{H}}=0$, where we subsequently quantify the induced current in $y$ direction. In Fig.~\ref{fig_1}(c), we show $\tilde{\sigma}_{xy}(t_{\text{H}})$ computed at the GS in the CI ($V=0$ and $1$) and CDW ($V=3$ and $4$) phases, respectively. We observe that the ground states with $C=1$ ($C=0$) have a nonzero (zero) Hall response, whose long-time values oscillate around the equilibrium ones obtained from the Kubo formula [Eq.~\eqref{eq:lin_resp}], which are denoted by the dashed lines. This indicates that the Hall response can be used to characterize the topological characteristics of the many-body wave function even in out-of-equilibrium, where an exact definition of the Chern number is elusive. We now proceed with the same protocol to compute $\tilde{\sigma}_{xy}(t_{\text{H}})$, but using the second excited state with $V=2.5$, $3$, $3.5$, $4$, as shown in Fig.~\ref{fig_1}(d). Although the Hall response decreases deep in the CDW phase, it is always finite, in direct agreement with the analysis of the corresponding Chern number of $|\psi_2\rangle$ in this regime. An important remark is that a quantization of the Hall response in these `equilibrium' settings can be obtained only for larger lattice sizes, in similarity to what occurs in non-interacting systems.

\begin{figure}[t]
\centering
\includegraphics[width=0.48\textwidth]{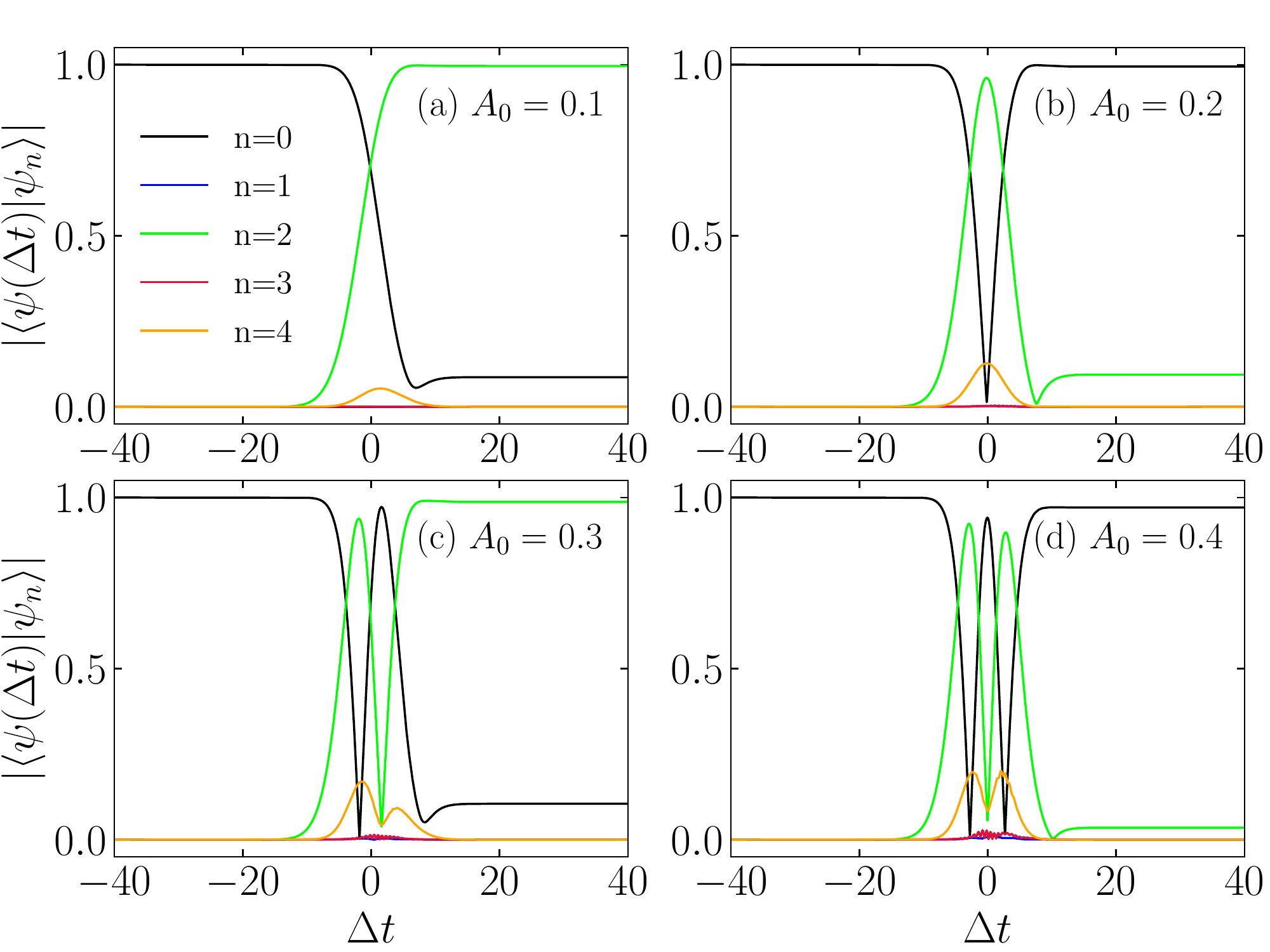}
\caption{(a) Overlaps between several eigenstates $|\psi_n\rangle$ and the time-dependent wave function $|\psi(\Delta t)\rangle$ under a pump with $A_0=0.1$ (a), $A_0=0.2$ (b), $A_0=0.3$ (c) and $A_0=0.4$ (d), respectively.
Pump parameters are $\omega_0=1.63$ and $t_d=4.0$; the interaction strength is set at $V=2.5$, within the CDW phase for the GS.}
\label{fig_2}
\end{figure}

\subsubsection{Pump dynamics}
\begin{figure}[t]
\centering
\includegraphics[width=0.48\textwidth]{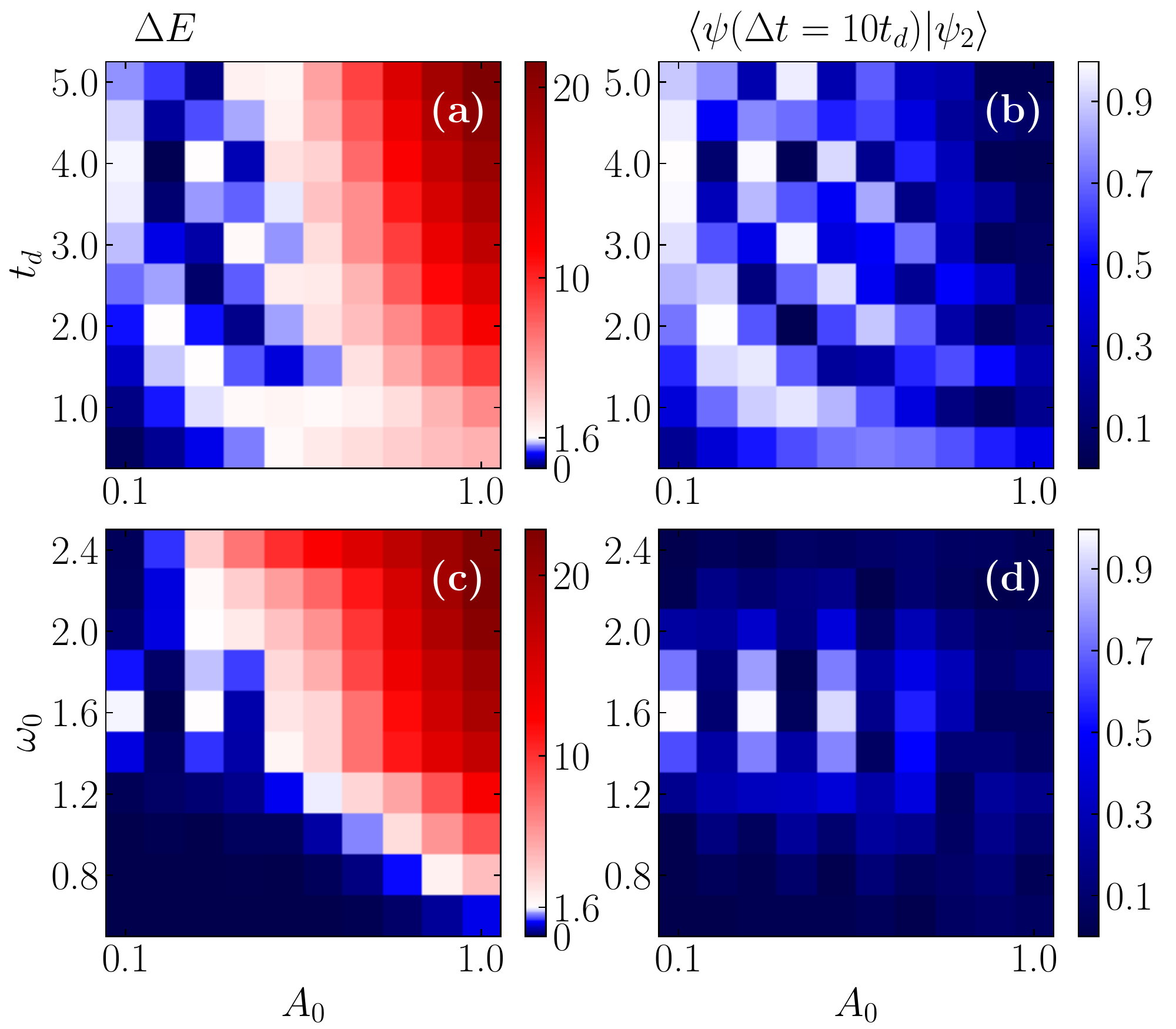}
\caption{Contour plots of the injected energy $\Delta E$ (a) and the overlap between $|\psi(\Delta t=10 t_d)\rangle$ and the eigenstate $|\psi_2\rangle$ (b) as a function of $A_0$ and $t_d$ with $\omega_0=1.63$. Contour plots of the injected energy $\Delta E$ (c) and the overlap $\langle\psi(\Delta t=10 t_d)|\psi_2\rangle$ (d) as a function of $A_0$ and $\omega_0$ with $t_d=4.0$. The color bars are adjusted in order to highlight the best conditions to resonantly excite the state $|\psi_2\rangle$, via the energy difference and the corresponding overlap with this state. In all cases, we fix the interaction at $V=2.5$.
}
\label{fig_3}
\end{figure}

Now that we have characterized the equilibrium scenario and used an out-of-equilibrium scheme to understand a possible dynamical Hall response, we turn focus to the possibility of generating nontrivial topological properties from a topologically trivial state, which includes two protocols: pump and quench. In the former, we emulate the ultrafast laser pulse by using the vector potential in Eqs.~\eqref{eq:Peierls} and \eqref{eq:vpotent}. By setting $V=2.5$, we investigate the overlaps between the time-dependent wave function $|\psi(\Delta t)\rangle$ and many eigenstates $|\psi_n\rangle$ of the equilibrium Hamiltonian, as shown in Fig.~\ref{fig_2}. The initial state $|\psi(\Delta t=-\infty)\rangle$ is the CDW ground state with $V=2.5$ and pumps with $\omega_0=1.63$ (that is, the pulse is made resonant with the energy difference $E_2-E_0\simeq1.63$), $t_d=4.0$ and four different $A_0$ are applied. The parameters' setting is chosen after a careful tuning, which will become more clear when describing Fig.~\ref{fig_3}.

In Figs.~\ref{fig_2}(a) and (c), with $A_0=0.1$ and $A_0=0.3$, respectively, we find that the overlap of $|\psi(\Delta t)\rangle$ with the $|\psi_2\rangle$ at long times after the pump can be larger than $0.98$, while the overlap with the initial CDW state is smaller than $0.1$, indicating a radical switch of the contribution of the two states in the time-dependent wave function. In contrast, the second excited state is mostly not excited in the case of $A_0=0.2$ and $A_0=0.4$, and the reason is due to the oscillation of the overlaps around the duration of the pulse, whose frequencies increases with growing $A_0$ (similar results with $V=4$ are shown in the Appendix~\ref{appendix1}). In addition, the evolution of the CDW structure factor for different $A_0$'s and associated discussion can be found in Appendix~\ref{appendix2}.

We conjecture that this oscillating behavior of the overlaps can be well explained via an analogy with a dynamical picture of the two-level Rabi model~\cite{Lu_2013}. Considering a two-level quantum system with $2\epsilon$ level spacing, coupled with a single-mode classical external field, the time-dependent Hamiltonian can be written as
\begin{eqnarray}
H_R(t)=\epsilon\sigma_z+{\sf g}(t)\sigma_x,
\label{eq:Rabi}
\end{eqnarray}
where $\sigma_a$ are Pauli matrices and ${\sf g}(t)=2g\cos(\omega t)$, which describes the coupling of the two-level system to an external field. The parameter $g$ here represents the strength of the external field and can be compared with the amplitude of the pump, i.e., $A_0$ in our case. When $\omega\approx2\epsilon$, which is the energy difference of the two levels, the rotating-wave approximation can be applied. The Rabi frequency, which characterizes the system's oscillation between the two levels, can be then expressed as
\begin{eqnarray}
\Omega_R=[(\epsilon-\omega/2)^2+g^2]^{1/2}.
\label{eq:Rabi}
\end{eqnarray}
As it increases with $g$, it can directly explain why larger $A_0$ leads to a larger oscillating frequency of overlaps in Fig.~\ref{fig_2}. Since the pulse is short-lived, the long-time overlaps can be either maximized at the original GS or the second excited state. Moreover, the pulse with higher $A_0$ provides sufficient energy to allow the participation of other eigenstates, and the two-level dynamics may eventually not work anymore.

To systematically investigate the influence of various pump parameters in these results, we show in Figs.~\ref{fig_3}(a) and (b) the contour plots of the injected energy $\Delta E = E(\Delta t\to \infty) - E_0$ and the overlap between $|\psi(\Delta t=10 t_d)\rangle$ and the eigenstate $|\psi_2\rangle$, as a function of $A_0$ and $t_d$. As before,  $\omega_0$ is set to be $1.63$, and we find that $A_0=0.1$ and $t_d=4.0$ is the optimal combination to increase the overlap with $|\psi_2\rangle$, which has been detailed in Fig.~\ref{fig_2}(a). We plot the same contour plots as a function of $A_0$ and $\omega_0$, in Figs.~\ref{fig_3}(c) and (d). Results in Fig.~\ref{fig_3}(d) confirm that $\omega_0\simeq1.63$ is the resonant frequency to excite $|\psi_2\rangle$. Besides, we observe that the overlaps have a staggered dependence on $A_0$, together with a staggered injected energy $\Delta E$, especially when $\omega_0\approx1.6$ and $t_d=4$. That is, the photoinduced increase of the overlap with eigenstate $|\psi_2\rangle$ strongly depends on $\Delta E$, and the settings of parameters that can predominantly excite $|\psi_2\rangle$ always occurs with injected energy $\Delta E\approx E_2 - E_0 \simeq 1.6$, which is just the energy $\hbar\omega_0$.

\begin{figure}[t]
\centering
\includegraphics[width=0.48\textwidth]{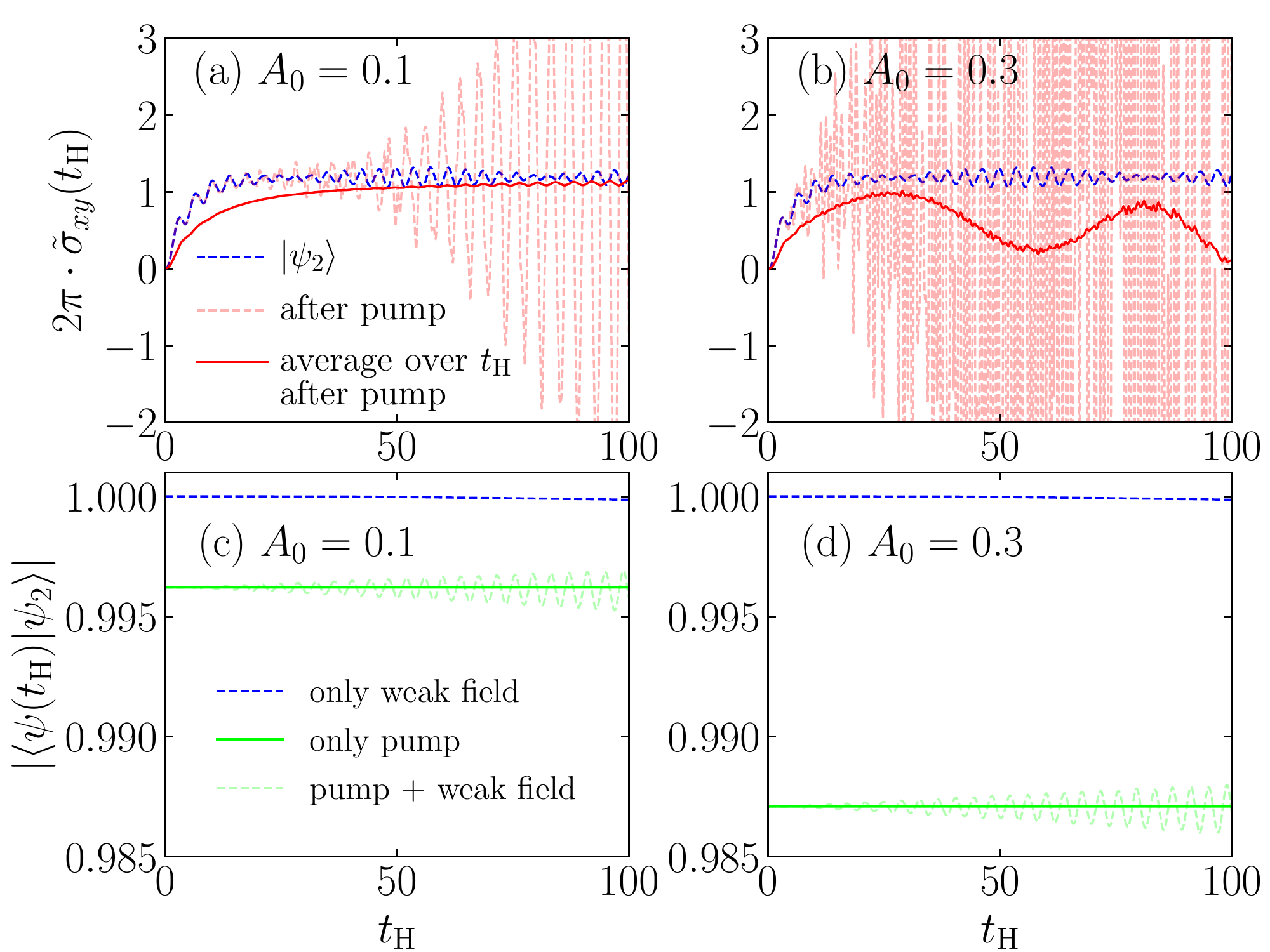}
\caption{The dynamical Hall response $\tilde{\sigma}_{xy}(t_{\text{H}})$ to the second excited state $|\psi_2\rangle$ with $V=2.5$ (blue dashed line) and to the time-dependent wave function after pump (red dashed lines) with $A_0=0.1$ (a) and $A_0=0.3$ (b). Red solid lines show the corresponding time-averages of $\tilde{\sigma}_{xy}(t_{\text{H}})$. In (c) and (d), we show overlaps between time-dependent wave function and the second excited state, $\langle\psi(t_{\rm H})|\psi_2\rangle$, after the weak electric field applied to $|\psi_2\rangle$ (blue dashed lines) and to the time-dependent wave function after pump (green dashed lines) with $A_0=0.1$ and $A_0=0.3$, respectively. Green solid lines show $\langle\psi(t_{\rm H})|\psi_2\rangle$ just after the pump without weak electric field applied. As explained in the text, the reference probe time $t_{\text{H}}=0$ locates at $\Delta t=40$. Parameters used: $V=2.5$, $\omega_0=1.63$ and $t_d=4.0$.}
\label{fig_4}
\end{figure}

To certify that a photoinduced topological phase transition takes place, we calculate the dynamical Hall response after pump with $A_0=0.1$ (probing from $\Delta t=40$, i.e, turning the probe electric field after a substantial time delayed from the central pump time), as shown in Fig.~\ref{fig_4}(a) by a red dashed line. For comparison, we also plot the `equilibrium' Hall response for state $|\psi_2\rangle$ [blue dashed line, similar data obtained in Fig.~\ref{fig_1}(d)], where we find that they are almost identical when $t_{\rm H} < 30$. At long times, continuous application of the weak electric field, however, leads the Hall response to display large oscillations but whose time-average value (red solid line) is still around the Hall response associated to $|\psi_2\rangle$. This points to the conclusion that the dynamical transition from topologically trivial to non-trivial state can be obtained after a well-tuned pump.

Now we explain the large oscillations of Hall response for the post-pump out-of-equilibrium state, which is not observed for equilibrium states in Fig.~\ref{fig_1}(c). The effect of weak electric field to the overlaps between the time-dependent wave functions and second excited state, $\langle\psi(t_{\rm H})|\psi_2\rangle$, is shown in Fig.~\ref{fig_4}(c) with $A_0 = 0.1$. If we just apply the weak field to $|\psi_2\rangle$, its influence on the overlap is rather small (see blue dashed line). However, if we apply the weak field to the post-pump time-dependent wave function (green dashed line), there is an apparent deviation and oscillation compared with that without weak electric field applied (green solid line). The reason is due to the time-dependent wave function never perfectly turning to $|\psi_2\rangle$ (the maximum overlap is 0.996), as expected from a unitary evolution that prevents $\langle \psi(t+\Delta t)|\psi(t)\rangle=0$~\cite{Ge21}. Even in the perfect resonant case, $|\psi(\Delta t\to\infty)\rangle$ is never an eigenstate of the Hamiltonian \eqref{eq:H}, in spite of abundant evidence showing the emergence of topological non-trivial properties. Moreover, the frequencies of oscillations on Hall response and overlap are identical. We then show the Hall response and overlaps with $A_0=0.3$ in Figs.~\ref{fig_4}(b) and (d), from which one can find that if the post-pump state being further away from $|\psi_2\rangle$ (0.987), it becomes hard to obtain a correct dynamical Hall response after $t_{\rm H}>10$, leading to an unstable time-average result.

\begin{figure}[t]
\centering
\includegraphics[width=0.48\textwidth]{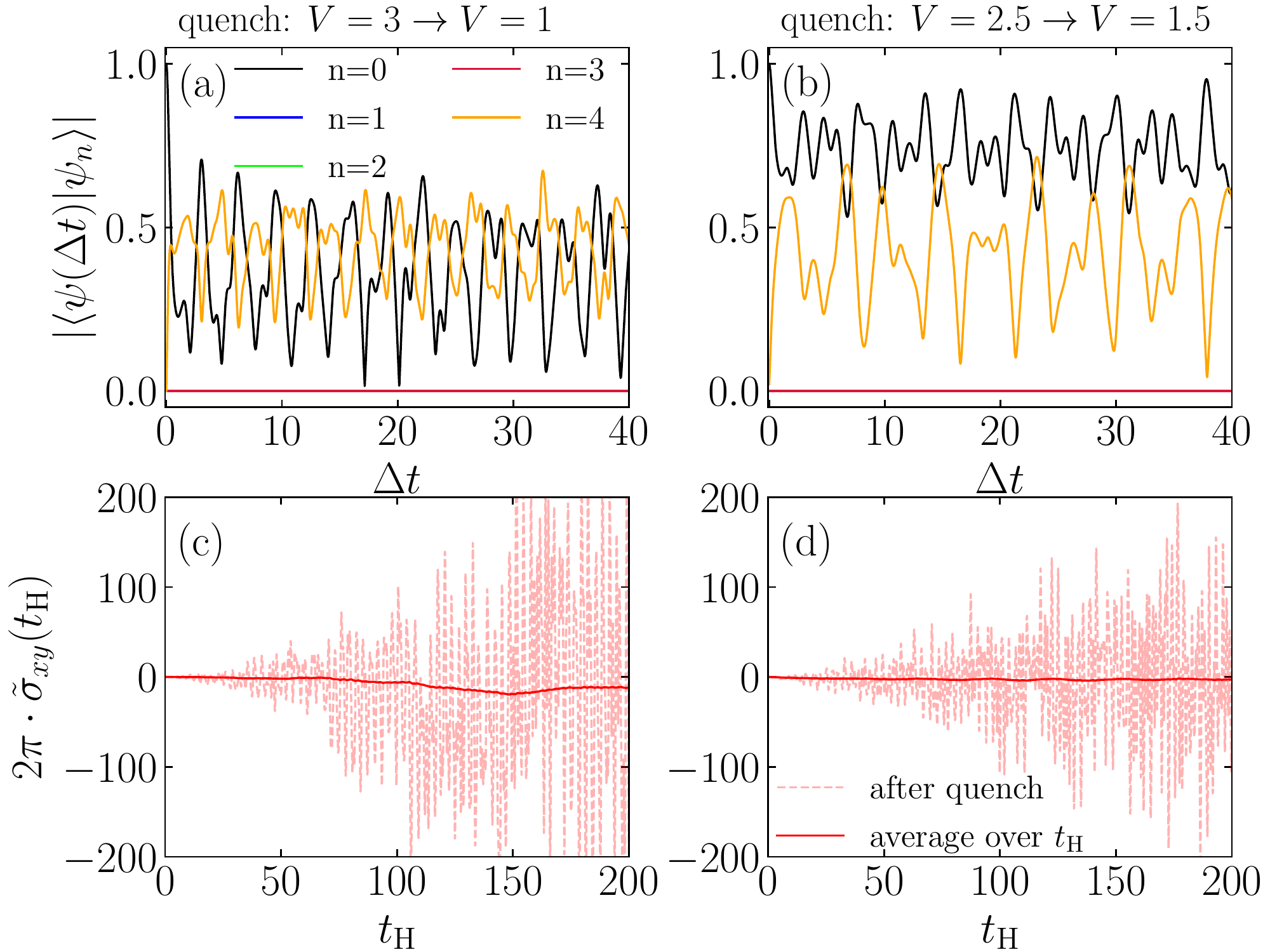}
\caption{Overlaps between several eigenstates $|\psi_n\rangle$'s and the time-dependent wave function $|\psi(\Delta t)\rangle$ after a quench from $V=3$ to $V=1$ (a) as well as from $V=2.5$ to $V=1.5$ (b). (c) and (d) show the Hall responses (red dashed lines) and their time average (red solid lines) to the nonequilibrium states after the corresponding quenches shown in (a) and (b), respectively.}
\label{fig_5}
\end{figure}

\subsubsection{Quench dynamics}
Finally, we contrast these results with more often used protocols to understand non-equilibrium topological transitions, that is, using a quench from a trivial to a non-trivial phase in the regime of parameters. As the `knob' in our model that destroys topological characteristics of the ground state is the interaction, we study a quench scenario from large interactions ($V>V_c$) to smaller ones ($V<V_c$). In Figs.~\ref{fig_5}(a) and (b), after the quenches $V=3.0\rightarrow1.0$ and $V=2.5\rightarrow1.5$, we calculate the overlaps between the time-dependent wave function and several eigenstates of the equilibrium Hamiltonian before quench. Although the overlap with the GS is substantially decreased in both cases, $|\psi_2\rangle$ is not excited. Instead, overlaps with the fourth excited state, whose Chern number can not be identified (owing to the small gaps to other states in the spectrum), increases considerably. In direct contrast to the pump case, the nonequilibrium state after quench is largely unstable, at least in a short-time scale. This can be inferred in Figs.~\ref{fig_5}(c) and (d), where we show the Hall responses after the quench (red dashed lines) and their time-average results (red solid line). We find no apparent features to characterize a potential topological phase transition.

\section{conclusions and discussion}\label{conclusion}

We investigated the interacting Haldane model that hosts a topological (Chern insulating) phase at small interaction strengths. In the strong coupling regime with a CDW ground state, we studied the topological features (including Chern number and Hall response) of an excited state, often neglected in previous investigations. For the first time, we allow this interacting system to evade equilibrium via the application of an external pump, instead of more common quench scenarios, and demonstrate that resonant targeting of topologically non-trivial excited states is possible, as long as they are separated from the continuum of the spectrum. This situation is precisely satisfied when a first-order phase transition occurs~\cite{Varney10,Varney11}, even in finite lattices. In contrast, if one focus on quench dynamics from topologically trivial to non-trivial cases, such an excited state cannot be stimulated, and no evident Hall response is found for the model under study.

\begin{figure}[t]
\centering
\includegraphics[width=0.48\textwidth]{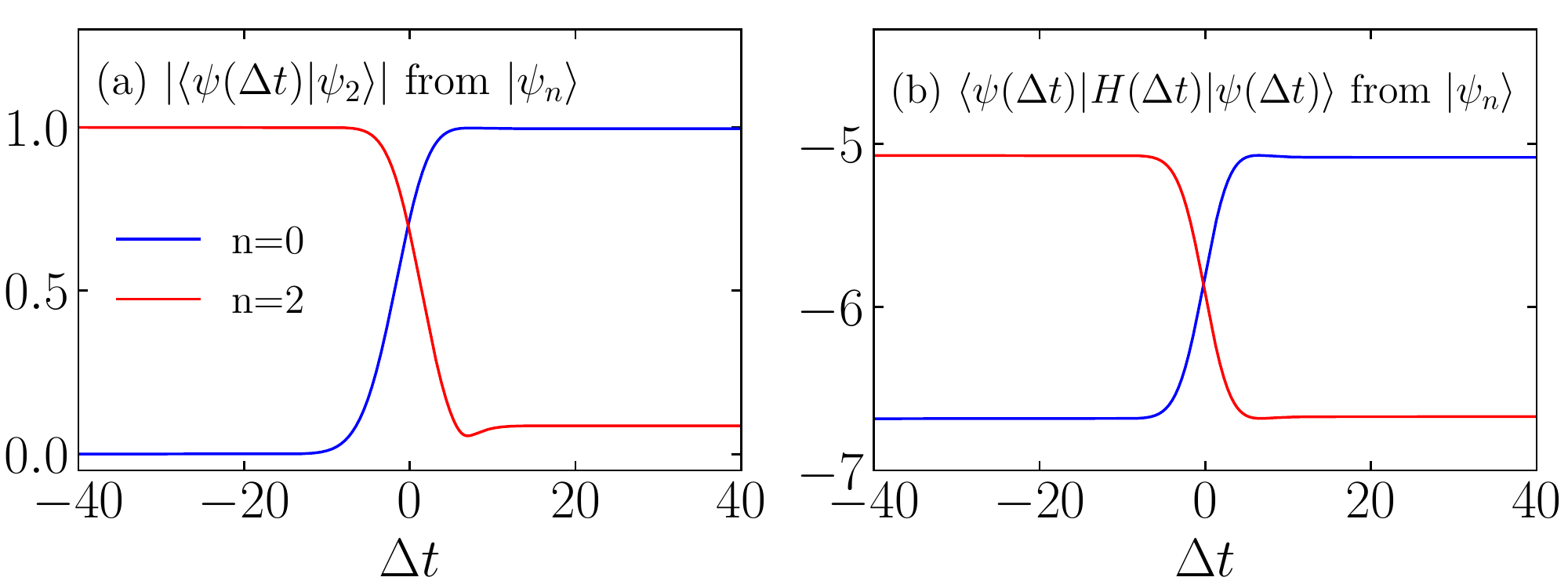}
\caption{Nonequilibrium dynamics after pump in the resonant conditions when starting from the ground state ($n=0$) and the second excited state ($n=2$) of the equilibrium Hamiltonian~\eqref{eq:H}. In (a) and (b), the overlaps with the $|\psi_2\rangle$ state and the instantaneous energies are shown for both different initializations; a characteristic dynamical level crossing occurs about the pump's central time. Here the pump parameters are the ones that optimize the targeting of the second excited state: $A_0=0.1$, $\omega_0 = 1.63$, and $t_d=4.0$, at interaction strength $V=2.5$.}
\label{fig_concl}
\end{figure}

To better understand how the dynamical topological transition is possible, we now compare the time-dependence of the overlaps and the system's instantaneous energy in the resonant conditions, when using both the original ground state $|\psi_0\rangle$, as well as the corresponding second excited eigenstate $|\psi_2\rangle$, as the initial states. By observing the overlaps with $|\psi_2\rangle$ [Fig.~\ref{fig_concl}(a)] and the total energy [Fig.~\ref{fig_concl}(b)] during the pump process, it becomes clear that a dynamical level crossing occurs for the two states, where they switch roles. This is at the core of our observation of induced Hall response. Given the first order character of the topological transition in equilibrium, such scenarios could be achieved in other models with engineered perturbations, attesting the generality of our results.

\begin{acknowledgments}
The authors acknowledges insightful discussions with H.~Lu and H.-Q.~Lin. C.~S.~acknowledges support from the National Natural Science Foundation of China (NSFC; Grants No.~12104229). P.~D.~S.~acknowledges the support from FCT through the Grant UID/CTM/04540/2019. R.~M.~acknowledges support from NSFC (Grants No. NSAF-U1930402, No.~11974039, No.~12050410263, and No.~12111530010). Computations were performed in the Tianhe-2JK at the Beijing Computational Science Research Center.
\end{acknowledgments}

\appendix
\section{Results of pump case with $V=4$}\label{appendix1}

\begin{figure}[t]
\centering
\includegraphics[width=0.48\textwidth]{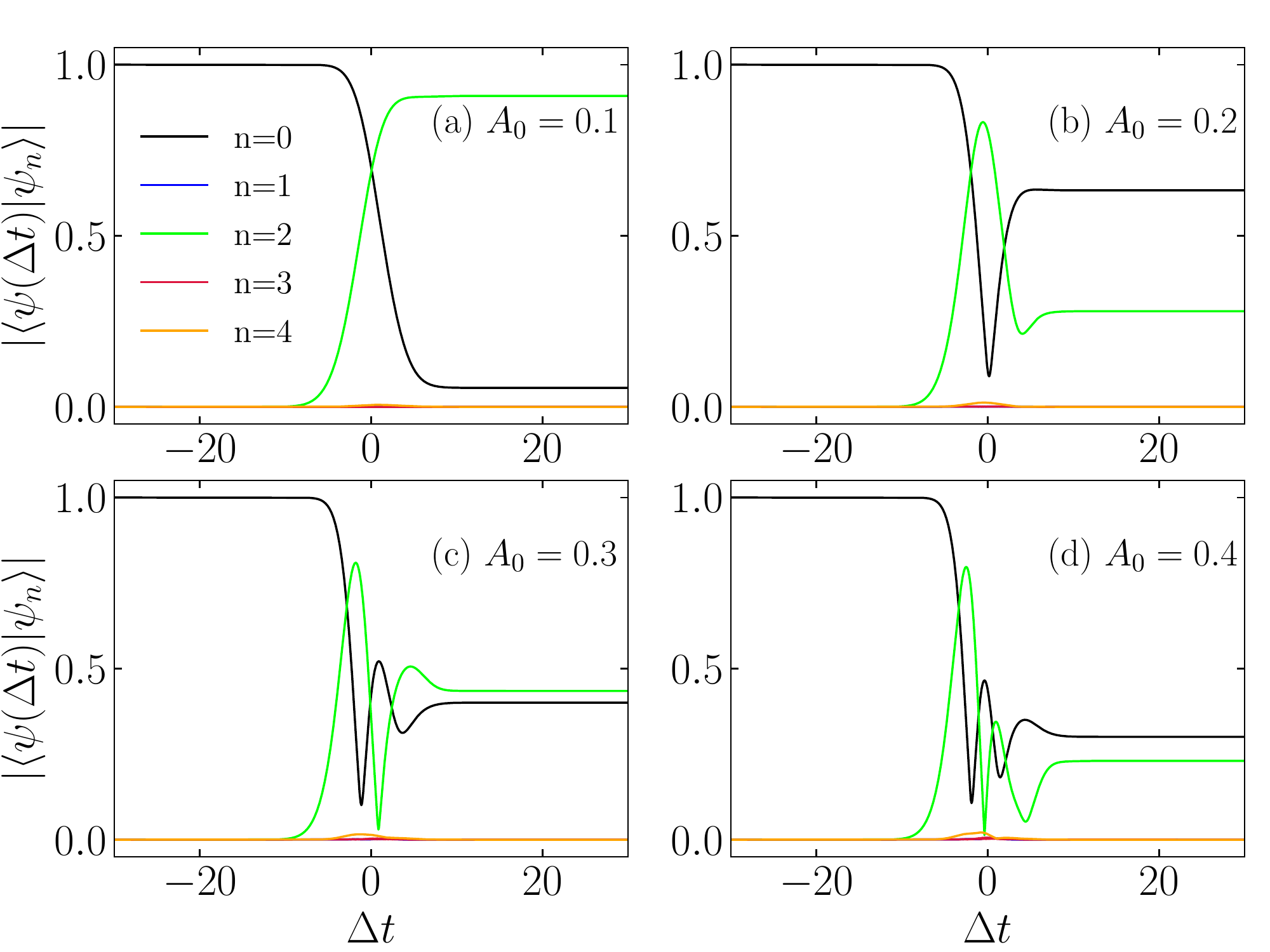}
\caption{(a) Overlaps between several eigenstates $|\psi_n\rangle$ and the time-dependent wave function $|\psi(\Delta t)\rangle$ under a pump with $A_0=0.1$ (a), $A_0=0.2$ (b), $A_0=0.3$ (c) and $A_0=0.4$ (d), respectively.
Pump parameters: $V=4.0$, $\omega_0=5.90$ and $t_d=3.0$.}
\label{fig_6}
\end{figure}

\begin{figure}
\centering
\includegraphics[width=0.48\textwidth]{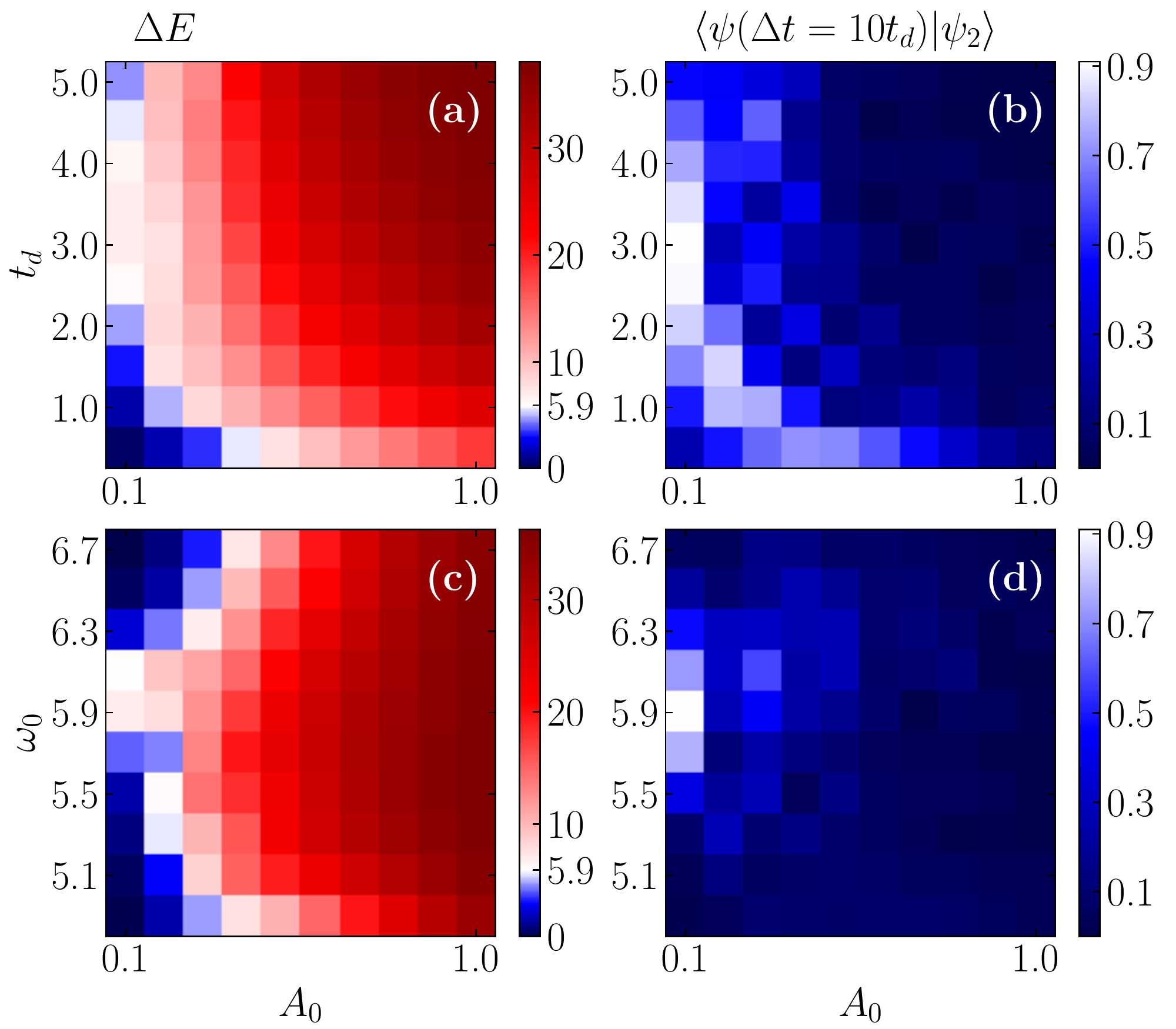}
\caption{Contour plots of the injected energy $\Delta E$ (a) and the overlap between $|\psi(\Delta t=10 t_d)\rangle$ and eigenstate $|\psi_2\rangle$ (b) as a function of $A_0$ and $t_d$ with $\omega_0=5.90$. Contour plots of the injected energy $\Delta E$ (c) and the overlap $\langle\psi(\Delta t=10 t_d)|\psi_2\rangle$ (d) as a function of $A_0$ and $\omega_0$ with $t_d=3.0$. Here the interaction is set at $V=4$.
}
\label{fig_7}
\end{figure}

In the main text, we focused on the interaction $V=2.5$ to demonstrate that the photoinduced topological phase transition can be realized. In this Appendix, we present similar results with $V=4$ to show that such nonequilibrium behavior is more general and still possible even when the system is further away from the quantum critical point $V_c\simeq 1.9$ in equilibrium~\cite{Varney11}.

Figure~\ref{fig_6} shows overlaps between the time-dependent wave function $|\psi(\Delta t)\rangle$ and several eigenstates $|\psi_n\rangle$ with $\omega_0=5.9$, $t_d=3.0$ and different $A_0$. The chosen set of parameters comes from the fine tuning in Fig.~\ref{fig_7}. In Fig.~\ref{fig_6}(a) with $A_0=0.1$, we find that the second excited state can be largely excited with a overlap of up to $0.91$. While for $A_0=0.3$, it is also excited in spite of the enhancement not being as large as for $A_0=0.1$. In Figs.~\ref{fig_7}(b) and (d), we show the contour plots of the overlaps between $|\psi(\Delta t=10 t_d)\rangle$ and the eigenstate $|\psi_2\rangle$, as a function of $A_0$ and $t_d$, as well as $A_0$ and $\omega_0$, respectively. The staggered enhancement of the overlaps can be seen with the parameters $t_d=3.0$ and $\omega_0=5.9$. As mentioned in the main text, a fundamental condition to produce resonant pumps is selecting the frequency $\omega_0$ such that it corresponds to the energy difference between the states one plans to promote excitations. Here in this case, $E_2-E_0\simeq 5.9$. Figures~\ref{fig_7}(a) and (c) show the contour plots of the injected energy $\Delta E$, where we can find that the injected energy of the optimal parameters to excite $|\psi_2\rangle$ is a little more than $5.9$. The reason is that more higher-energy eigenstates participate in the pump dynamics. These results indicate that even away from the phase transition point, it is still possible to induce a topological phase transition via a well-tuned laser pulse.

\begin{figure}[b]
\centering
\includegraphics[width=0.3\textwidth]{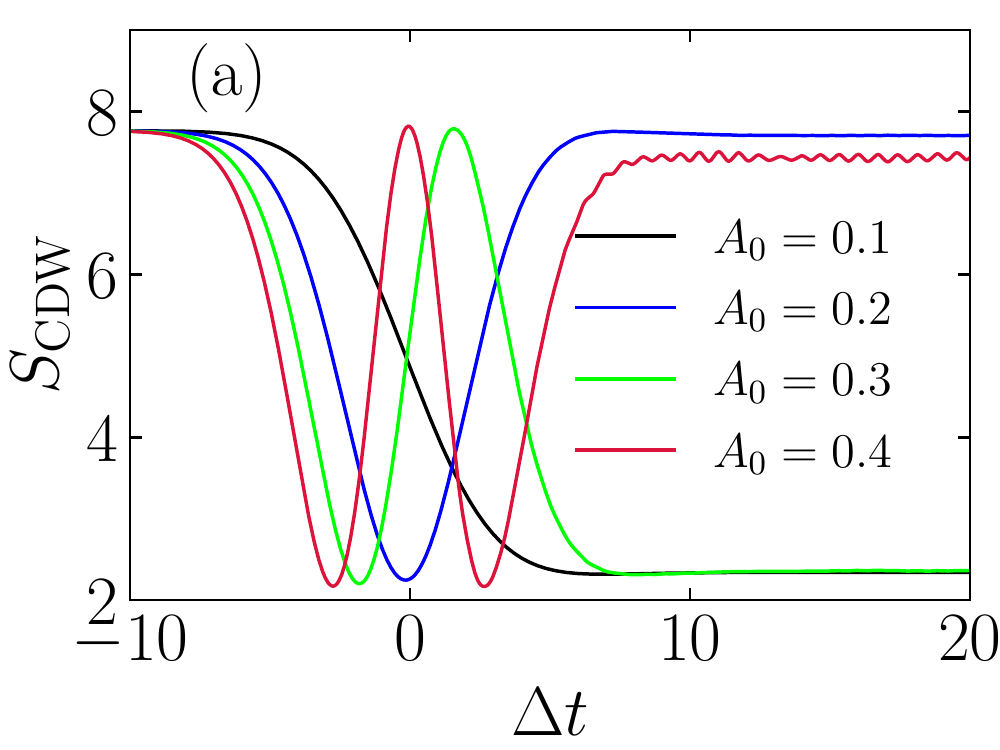}
\caption{Time-dependent CDW structure factor with the pump amplitude $A_0=0.1$, $0.2$, $0.3$ and $0.4$, respectively. Parameters used: $V=2.5$, $\omega_0=1.63$ and $t_d=4.0$.}
\label{fig_8}
\end{figure}

\section{Time-dependent CDW structure factor}\label{appendix2}

Figure~\ref{fig_8} displays the time-dependent CDW structure factor with the resonant pump parameters $\omega_0=1.63$ and $t_d=4.0$, and interaction $V=2.5$. The changes of the $S_{\rm CDW}$ for different $A_0$ are precisely consistent with that of overlaps between time-dependent wave functions and the GS in Fig.~\ref{fig_2}. When $A_0 = 0.1$ or $0.3$, the collapses of CDW state can be found after pump, since in such cases $|\psi(t)\rangle$ has a large overlap with a state that does not display charge order.


\begin{thebibliography}{55}%
\makeatletter
\providecommand \@ifxundefined [1]{%
 \@ifx{#1\undefined}
}%
\providecommand \@ifnum [1]{%
 \ifnum #1\expandafter \@firstoftwo
 \else \expandafter \@secondoftwo
 \fi
}%
\providecommand \@ifx [1]{%
 \ifx #1\expandafter \@firstoftwo
 \else \expandafter \@secondoftwo
 \fi
}%
\providecommand \natexlab [1]{#1}%
\providecommand \enquote  [1]{``#1''}%
\providecommand \bibnamefont  [1]{#1}%
\providecommand \bibfnamefont [1]{#1}%
\providecommand \citenamefont [1]{#1}%
\providecommand \href@noop [0]{\@secondoftwo}%
\providecommand \href [0]{\begingroup \@sanitize@url \@href}%
\providecommand \@href[1]{\@@startlink{#1}\@@href}%
\providecommand \@@href[1]{\endgroup#1\@@endlink}%
\providecommand \@sanitize@url [0]{\catcode `\\12\catcode `\$12\catcode
  `\&12\catcode `\#12\catcode `\^12\catcode `\_12\catcode `\%12\relax}%
\providecommand \@@startlink[1]{}%
\providecommand \@@endlink[0]{}%
\providecommand \url  [0]{\begingroup\@sanitize@url \@url }%
\providecommand \@url [1]{\endgroup\@href {#1}{\urlprefix }}%
\providecommand \urlprefix  [0]{URL }%
\providecommand \Eprint [0]{\href }%
\providecommand \doibase [0]{http://dx.doi.org/}%
\providecommand \selectlanguage [0]{\@gobble}%
\providecommand \bibinfo  [0]{\@secondoftwo}%
\providecommand \bibfield  [0]{\@secondoftwo}%
\providecommand \translation [1]{[#1]}%
\providecommand \BibitemOpen [0]{}%
\providecommand \bibitemStop [0]{}%
\providecommand \bibitemNoStop [0]{.\EOS\space}%
\providecommand \EOS [0]{\spacefactor3000\relax}%
\providecommand \BibitemShut  [1]{\csname bibitem#1\endcsname}%
\let\auto@bib@innerbib\@empty
\bibitem [{\citenamefont {Chiu}\ \emph {et~al.}(2016)\citenamefont {Chiu},
  \citenamefont {Teo}, \citenamefont {Schnyder},\ and\ \citenamefont
  {Ryu}}]{Chiu16}%
  \BibitemOpen
  \bibfield  {author} {\bibinfo {author} {\bibfnamefont {C.-K.}\ \bibnamefont
  {Chiu}}, \bibinfo {author} {\bibfnamefont {J.~C.~Y.}\ \bibnamefont {Teo}},
  \bibinfo {author} {\bibfnamefont {A.~P.}\ \bibnamefont {Schnyder}}, \ and\
  \bibinfo {author} {\bibfnamefont {S.}~\bibnamefont {Ryu}},\ }\href {\doibase
  10.1103/RevModPhys.88.035005} {\bibfield  {journal} {\bibinfo  {journal}
  {Rev. Mod. Phys.}\ }\textbf {\bibinfo {volume} {88}},\ \bibinfo {pages}
  {035005} (\bibinfo {year} {2016})}\BibitemShut {NoStop}%
\bibitem [{\citenamefont {Kruthoff}\ \emph {et~al.}(2017)\citenamefont
  {Kruthoff}, \citenamefont {de~Boer}, \citenamefont {van Wezel}, \citenamefont
  {Kane},\ and\ \citenamefont {Slager}}]{Kruthoff17}%
  \BibitemOpen
  \bibfield  {author} {\bibinfo {author} {\bibfnamefont {J.}~\bibnamefont
  {Kruthoff}}, \bibinfo {author} {\bibfnamefont {J.}~\bibnamefont {de~Boer}},
  \bibinfo {author} {\bibfnamefont {J.}~\bibnamefont {van Wezel}}, \bibinfo
  {author} {\bibfnamefont {C.~L.}\ \bibnamefont {Kane}}, \ and\ \bibinfo
  {author} {\bibfnamefont {R.-J.}\ \bibnamefont {Slager}},\ }\href {\doibase
  10.1103/PhysRevX.7.041069} {\bibfield  {journal} {\bibinfo  {journal} {Phys.
  Rev. X}\ }\textbf {\bibinfo {volume} {7}},\ \bibinfo {pages} {041069}
  (\bibinfo {year} {2017})}\BibitemShut {NoStop}%
\bibitem [{\citenamefont {Zhang}\ \emph {et~al.}(2019)\citenamefont {Zhang},
  \citenamefont {Jiang}, \citenamefont {Song}, \citenamefont {Huang},
  \citenamefont {He}, \citenamefont {Fang}, \citenamefont {Weng},\ and\
  \citenamefont {Fang}}]{Zhang19}%
  \BibitemOpen
  \bibfield  {author} {\bibinfo {author} {\bibfnamefont {T.}~\bibnamefont
  {Zhang}}, \bibinfo {author} {\bibfnamefont {Y.}~\bibnamefont {Jiang}},
  \bibinfo {author} {\bibfnamefont {Z.}~\bibnamefont {Song}}, \bibinfo {author}
  {\bibfnamefont {H.}~\bibnamefont {Huang}}, \bibinfo {author} {\bibfnamefont
  {Y.}~\bibnamefont {He}}, \bibinfo {author} {\bibfnamefont {Z.}~\bibnamefont
  {Fang}}, \bibinfo {author} {\bibfnamefont {H.}~\bibnamefont {Weng}}, \ and\
  \bibinfo {author} {\bibfnamefont {C.}~\bibnamefont {Fang}},\ }\href {\doibase
  10.1038/s41586-019-0944-6} {\bibfield  {journal} {\bibinfo  {journal}
  {Nature}\ }\textbf {\bibinfo {volume} {566}},\ \bibinfo {pages} {475}
  (\bibinfo {year} {2019})}\BibitemShut {NoStop}%
\bibitem [{\citenamefont {Vergniory}\ \emph {et~al.}(2019)\citenamefont
  {Vergniory}, \citenamefont {Elcoro}, \citenamefont {Felser}, \citenamefont
  {Regnault}, \citenamefont {Bernevig},\ and\ \citenamefont
  {Wang}}]{Vergniory19}%
  \BibitemOpen
  \bibfield  {author} {\bibinfo {author} {\bibfnamefont {M.~G.}\ \bibnamefont
  {Vergniory}}, \bibinfo {author} {\bibfnamefont {L.}~\bibnamefont {Elcoro}},
  \bibinfo {author} {\bibfnamefont {C.}~\bibnamefont {Felser}}, \bibinfo
  {author} {\bibfnamefont {N.}~\bibnamefont {Regnault}}, \bibinfo {author}
  {\bibfnamefont {B.~A.}\ \bibnamefont {Bernevig}}, \ and\ \bibinfo {author}
  {\bibfnamefont {Z.}~\bibnamefont {Wang}},\ }\href {\doibase
  10.1038/s41586-019-0954-4} {\bibfield  {journal} {\bibinfo  {journal}
  {Nature}\ }\textbf {\bibinfo {volume} {566}},\ \bibinfo {pages} {480}
  (\bibinfo {year} {2019})}\BibitemShut {NoStop}%
\bibitem [{\citenamefont {Tang}\ \emph {et~al.}(2019)\citenamefont {Tang},
  \citenamefont {Po}, \citenamefont {Vishwanath},\ and\ \citenamefont
  {Wan}}]{Tang19}%
  \BibitemOpen
  \bibfield  {author} {\bibinfo {author} {\bibfnamefont {F.}~\bibnamefont
  {Tang}}, \bibinfo {author} {\bibfnamefont {H.~C.}\ \bibnamefont {Po}},
  \bibinfo {author} {\bibfnamefont {A.}~\bibnamefont {Vishwanath}}, \ and\
  \bibinfo {author} {\bibfnamefont {X.}~\bibnamefont {Wan}},\ }\href {\doibase
  10.1038/s41586-019-0937-5} {\bibfield  {journal} {\bibinfo  {journal}
  {Nature}\ }\textbf {\bibinfo {volume} {566}},\ \bibinfo {pages} {486}
  (\bibinfo {year} {2019})}\BibitemShut {NoStop}%
\bibitem [{\citenamefont {Bansil}\ \emph {et~al.}(2016)\citenamefont {Bansil},
  \citenamefont {Lin},\ and\ \citenamefont {Das}}]{Bansil16}%
  \BibitemOpen
  \bibfield  {author} {\bibinfo {author} {\bibfnamefont {A.}~\bibnamefont
  {Bansil}}, \bibinfo {author} {\bibfnamefont {H.}~\bibnamefont {Lin}}, \ and\
  \bibinfo {author} {\bibfnamefont {T.}~\bibnamefont {Das}},\ }\href {\doibase
  10.1103/RevModPhys.88.021004} {\bibfield  {journal} {\bibinfo  {journal}
  {Rev. Mod. Phys.}\ }\textbf {\bibinfo {volume} {88}},\ \bibinfo {pages}
  {021004} (\bibinfo {year} {2016})}\BibitemShut {NoStop}%
\bibitem [{\citenamefont {Jotzu}\ \emph {et~al.}(2014)\citenamefont {Jotzu},
  \citenamefont {Messer}, \citenamefont {Desbuquois}, \citenamefont {Lebrat},
  \citenamefont {Uehlinger}, \citenamefont {Greif},\ and\ \citenamefont
  {Esslinger}}]{Jotzu14}%
  \BibitemOpen
  \bibfield  {author} {\bibinfo {author} {\bibfnamefont {G.}~\bibnamefont
  {Jotzu}}, \bibinfo {author} {\bibfnamefont {M.}~\bibnamefont {Messer}},
  \bibinfo {author} {\bibfnamefont {R.}~\bibnamefont {Desbuquois}}, \bibinfo
  {author} {\bibfnamefont {M.}~\bibnamefont {Lebrat}}, \bibinfo {author}
  {\bibfnamefont {T.}~\bibnamefont {Uehlinger}}, \bibinfo {author}
  {\bibfnamefont {D.}~\bibnamefont {Greif}}, \ and\ \bibinfo {author}
  {\bibfnamefont {T.}~\bibnamefont {Esslinger}},\ }\href {\doibase
  10.1038/nature13915} {\bibfield  {journal} {\bibinfo  {journal} {Nature}\
  }\textbf {\bibinfo {volume} {515}},\ \bibinfo {pages} {237} (\bibinfo {year}
  {2014})}\BibitemShut {NoStop}%
\bibitem [{\citenamefont {Aidelsburger}\ \emph {et~al.}(2015)\citenamefont
  {Aidelsburger}, \citenamefont {Lohse}, \citenamefont {Schweizer},
  \citenamefont {Atala}, \citenamefont {Barreiro}, \citenamefont
  {Nascimb{\`e}ne}, \citenamefont {Cooper}, \citenamefont {Bloch},\ and\
  \citenamefont {Goldman}}]{Aidelsburger15}%
  \BibitemOpen
  \bibfield  {author} {\bibinfo {author} {\bibfnamefont {M.}~\bibnamefont
  {Aidelsburger}}, \bibinfo {author} {\bibfnamefont {M.}~\bibnamefont {Lohse}},
  \bibinfo {author} {\bibfnamefont {C.}~\bibnamefont {Schweizer}}, \bibinfo
  {author} {\bibfnamefont {M.}~\bibnamefont {Atala}}, \bibinfo {author}
  {\bibfnamefont {J.~T.}\ \bibnamefont {Barreiro}}, \bibinfo {author}
  {\bibfnamefont {S.}~\bibnamefont {Nascimb{\`e}ne}}, \bibinfo {author}
  {\bibfnamefont {N.~R.}\ \bibnamefont {Cooper}}, \bibinfo {author}
  {\bibfnamefont {I.}~\bibnamefont {Bloch}}, \ and\ \bibinfo {author}
  {\bibfnamefont {N.}~\bibnamefont {Goldman}},\ }\href {\doibase
  10.1038/nphys3171} {\bibfield  {journal} {\bibinfo  {journal} {Nat. Phys.}\
  }\textbf {\bibinfo {volume} {11}},\ \bibinfo {pages} {162} (\bibinfo {year}
  {2015})}\BibitemShut {NoStop}%
\bibitem [{\citenamefont {Chen}\ \emph {et~al.}(2010)\citenamefont {Chen},
  \citenamefont {Gu},\ and\ \citenamefont {Wen}}]{Chen10}%
  \BibitemOpen
  \bibfield  {author} {\bibinfo {author} {\bibfnamefont {X.}~\bibnamefont
  {Chen}}, \bibinfo {author} {\bibfnamefont {Z.-C.}\ \bibnamefont {Gu}}, \ and\
  \bibinfo {author} {\bibfnamefont {X.-G.}\ \bibnamefont {Wen}},\ }\href
  {\doibase 10.1103/PhysRevB.82.155138} {\bibfield  {journal} {\bibinfo
  {journal} {Phys. Rev. B}\ }\textbf {\bibinfo {volume} {82}},\ \bibinfo
  {pages} {155138} (\bibinfo {year} {2010})}\BibitemShut {NoStop}%
\bibitem [{\citenamefont {Foster}\ \emph {et~al.}(2013)\citenamefont {Foster},
  \citenamefont {Dzero}, \citenamefont {Gurarie},\ and\ \citenamefont
  {Yuzbashyan}}]{Foster13}%
  \BibitemOpen
  \bibfield  {author} {\bibinfo {author} {\bibfnamefont {M.~S.}\ \bibnamefont
  {Foster}}, \bibinfo {author} {\bibfnamefont {M.}~\bibnamefont {Dzero}},
  \bibinfo {author} {\bibfnamefont {V.}~\bibnamefont {Gurarie}}, \ and\
  \bibinfo {author} {\bibfnamefont {E.~A.}\ \bibnamefont {Yuzbashyan}},\ }\href
  {\doibase 10.1103/PhysRevB.88.104511} {\bibfield  {journal} {\bibinfo
  {journal} {Phys. Rev. B}\ }\textbf {\bibinfo {volume} {88}},\ \bibinfo
  {pages} {104511} (\bibinfo {year} {2013})}\BibitemShut {NoStop}%
\bibitem [{\citenamefont {Sacramento}(2014)}]{Sacramento14}%
  \BibitemOpen
  \bibfield  {author} {\bibinfo {author} {\bibfnamefont {P.~D.}\ \bibnamefont
  {Sacramento}},\ }\href {\doibase 10.1103/PhysRevE.90.032138} {\bibfield
  {journal} {\bibinfo  {journal} {Phys. Rev. E}\ }\textbf {\bibinfo {volume}
  {90}},\ \bibinfo {pages} {032138} (\bibinfo {year} {2014})}\BibitemShut
  {NoStop}%
\bibitem [{\citenamefont {D'Alessio}\ and\ \citenamefont
  {Rigol}(2015)}]{Alessio15}%
  \BibitemOpen
  \bibfield  {author} {\bibinfo {author} {\bibfnamefont {L.}~\bibnamefont
  {D'Alessio}}\ and\ \bibinfo {author} {\bibfnamefont {M.}~\bibnamefont
  {Rigol}},\ }\href {\doibase 10.1038/ncomms9336} {\bibfield  {journal}
  {\bibinfo  {journal} {Nature Communications}\ }\textbf {\bibinfo {volume}
  {6}},\ \bibinfo {pages} {8336} (\bibinfo {year} {2015})}\BibitemShut
  {NoStop}%
\bibitem [{\citenamefont {Sacramento}(2016)}]{Sacramento16}%
  \BibitemOpen
  \bibfield  {author} {\bibinfo {author} {\bibfnamefont {P.~D.}\ \bibnamefont
  {Sacramento}},\ }\href {\doibase 10.1103/PhysRevE.93.062117} {\bibfield
  {journal} {\bibinfo  {journal} {Phys. Rev. E}\ }\textbf {\bibinfo {volume}
  {93}},\ \bibinfo {pages} {062117} (\bibinfo {year} {2016})}\BibitemShut
  {NoStop}%
\bibitem [{\citenamefont {Caio}\ \emph {et~al.}(2016)\citenamefont {Caio},
  \citenamefont {Cooper},\ and\ \citenamefont {Bhaseen}}]{Caio16}%
  \BibitemOpen
  \bibfield  {author} {\bibinfo {author} {\bibfnamefont {M.~D.}\ \bibnamefont
  {Caio}}, \bibinfo {author} {\bibfnamefont {N.~R.}\ \bibnamefont {Cooper}}, \
  and\ \bibinfo {author} {\bibfnamefont {M.~J.}\ \bibnamefont {Bhaseen}},\
  }\href {\doibase 10.1103/PhysRevB.94.155104} {\bibfield  {journal} {\bibinfo
  {journal} {Phys. Rev. B}\ }\textbf {\bibinfo {volume} {94}},\ \bibinfo
  {pages} {155104} (\bibinfo {year} {2016})}\BibitemShut {NoStop}%
\bibitem [{\citenamefont {Hu}\ \emph {et~al.}(2016)\citenamefont {Hu},
  \citenamefont {Zoller},\ and\ \citenamefont {Budich}}]{Ying16}%
  \BibitemOpen
  \bibfield  {author} {\bibinfo {author} {\bibfnamefont {Y.}~\bibnamefont
  {Hu}}, \bibinfo {author} {\bibfnamefont {P.}~\bibnamefont {Zoller}}, \ and\
  \bibinfo {author} {\bibfnamefont {J.~C.}\ \bibnamefont {Budich}},\ }\href
  {\doibase 10.1103/PhysRevLett.117.126803} {\bibfield  {journal} {\bibinfo
  {journal} {Phys. Rev. Lett.}\ }\textbf {\bibinfo {volume} {117}},\ \bibinfo
  {pages} {126803} (\bibinfo {year} {2016})}\BibitemShut {NoStop}%
\bibitem [{\citenamefont {Kruckenhauser}\ and\ \citenamefont
  {Budich}(2018)}]{Kruckenhauser18}%
  \BibitemOpen
  \bibfield  {author} {\bibinfo {author} {\bibfnamefont {A.}~\bibnamefont
  {Kruckenhauser}}\ and\ \bibinfo {author} {\bibfnamefont {J.~C.}\ \bibnamefont
  {Budich}},\ }\href {\doibase 10.1103/PhysRevB.98.195124} {\bibfield
  {journal} {\bibinfo  {journal} {Phys. Rev. B}\ }\textbf {\bibinfo {volume}
  {98}},\ \bibinfo {pages} {195124} (\bibinfo {year} {2018})}\BibitemShut
  {NoStop}%
\bibitem [{\citenamefont {Sch\"uler}\ \emph {et~al.}(2019)\citenamefont
  {Sch\"uler}, \citenamefont {Budich},\ and\ \citenamefont
  {Werner}}]{Michael19}%
  \BibitemOpen
  \bibfield  {author} {\bibinfo {author} {\bibfnamefont {M.}~\bibnamefont
  {Sch\"uler}}, \bibinfo {author} {\bibfnamefont {J.~C.}\ \bibnamefont
  {Budich}}, \ and\ \bibinfo {author} {\bibfnamefont {P.}~\bibnamefont
  {Werner}},\ }\href {\doibase 10.1103/PhysRevB.100.041101} {\bibfield
  {journal} {\bibinfo  {journal} {Phys. Rev. B}\ }\textbf {\bibinfo {volume}
  {100}},\ \bibinfo {pages} {041101(R)} (\bibinfo {year} {2019})}\BibitemShut
  {NoStop}%
\bibitem [{\citenamefont {Thouless}\ \emph {et~al.}(1982)\citenamefont
  {Thouless}, \citenamefont {Kohmoto}, \citenamefont {Nightingale},\ and\
  \citenamefont {denNijs}}]{Thouless82}%
  \BibitemOpen
  \bibfield  {author} {\bibinfo {author} {\bibfnamefont {D.~J.}\ \bibnamefont
  {Thouless}}, \bibinfo {author} {\bibfnamefont {M.}~\bibnamefont {Kohmoto}},
  \bibinfo {author} {\bibfnamefont {M.~P.}\ \bibnamefont {Nightingale}}, \ and\
  \bibinfo {author} {\bibfnamefont {M.}~\bibnamefont {denNijs}},\ }\href
  {\doibase 10.1103/PhysRevLett.49.405} {\bibfield  {journal} {\bibinfo
  {journal} {Phys. Rev. Lett.}\ }\textbf {\bibinfo {volume} {49}},\ \bibinfo
  {pages} {405} (\bibinfo {year} {1982})}\BibitemShut {NoStop}%
\bibitem [{\citenamefont {Wilson}\ \emph {et~al.}(2016)\citenamefont {Wilson},
  \citenamefont {Song},\ and\ \citenamefont {Refael}}]{Wilson16}%
  \BibitemOpen
  \bibfield  {author} {\bibinfo {author} {\bibfnamefont {J.~H.}\ \bibnamefont
  {Wilson}}, \bibinfo {author} {\bibfnamefont {J.~C.~W.}\ \bibnamefont {Song}},
  \ and\ \bibinfo {author} {\bibfnamefont {G.}~\bibnamefont {Refael}},\ }\href
  {\doibase 10.1103/PhysRevLett.117.235302} {\bibfield  {journal} {\bibinfo
  {journal} {Phys. Rev. Lett.}\ }\textbf {\bibinfo {volume} {117}},\ \bibinfo
  {pages} {235302} (\bibinfo {year} {2016})}\BibitemShut {NoStop}%
\bibitem [{\citenamefont {Ge}\ and\ \citenamefont {Rigol}(2021)}]{Ge21}%
  \BibitemOpen
  \bibfield  {author} {\bibinfo {author} {\bibfnamefont {Y.}~\bibnamefont
  {Ge}}\ and\ \bibinfo {author} {\bibfnamefont {M.}~\bibnamefont {Rigol}},\
  }\href {\doibase 10.1103/PhysRevA.103.013314} {\bibfield  {journal} {\bibinfo
   {journal} {Phys. Rev. A}\ }\textbf {\bibinfo {volume} {103}},\ \bibinfo
  {pages} {013314} (\bibinfo {year} {2021})}\BibitemShut {NoStop}%
\bibitem [{\citenamefont {Peralta~Gavensky}\ \emph {et~al.}(2018)\citenamefont
  {Peralta~Gavensky}, \citenamefont {Usaj},\ and\ \citenamefont
  {Balseiro}}]{Peralta18}%
  \BibitemOpen
  \bibfield  {author} {\bibinfo {author} {\bibfnamefont {L.}~\bibnamefont
  {Peralta~Gavensky}}, \bibinfo {author} {\bibfnamefont {G.}~\bibnamefont
  {Usaj}}, \ and\ \bibinfo {author} {\bibfnamefont {C.~A.}\ \bibnamefont
  {Balseiro}},\ }\href {\doibase 10.1103/PhysRevB.98.165414} {\bibfield
  {journal} {\bibinfo  {journal} {Phys. Rev. B}\ }\textbf {\bibinfo {volume}
  {98}},\ \bibinfo {pages} {165414} (\bibinfo {year} {2018})}\BibitemShut
  {NoStop}%
\bibitem [{\citenamefont {Varney}\ \emph {et~al.}(2010)\citenamefont {Varney},
  \citenamefont {Sun}, \citenamefont {Rigol},\ and\ \citenamefont
  {Galitski}}]{Varney10}%
  \BibitemOpen
  \bibfield  {author} {\bibinfo {author} {\bibfnamefont {C.~N.}\ \bibnamefont
  {Varney}}, \bibinfo {author} {\bibfnamefont {K.}~\bibnamefont {Sun}},
  \bibinfo {author} {\bibfnamefont {M.}~\bibnamefont {Rigol}}, \ and\ \bibinfo
  {author} {\bibfnamefont {V.}~\bibnamefont {Galitski}},\ }\href {\doibase
  10.1103/PhysRevB.82.115125} {\bibfield  {journal} {\bibinfo  {journal} {Phys.
  Rev. B}\ }\textbf {\bibinfo {volume} {82}},\ \bibinfo {pages} {115125}
  (\bibinfo {year} {2010})}\BibitemShut {NoStop}%
\bibitem [{\citenamefont {Varney}\ \emph {et~al.}(2011)\citenamefont {Varney},
  \citenamefont {Sun}, \citenamefont {Rigol},\ and\ \citenamefont
  {Galitski}}]{Varney11}%
  \BibitemOpen
  \bibfield  {author} {\bibinfo {author} {\bibfnamefont {C.~N.}\ \bibnamefont
  {Varney}}, \bibinfo {author} {\bibfnamefont {K.}~\bibnamefont {Sun}},
  \bibinfo {author} {\bibfnamefont {M.}~\bibnamefont {Rigol}}, \ and\ \bibinfo
  {author} {\bibfnamefont {V.}~\bibnamefont {Galitski}},\ }\href {\doibase
  10.1103/PhysRevB.84.241105} {\bibfield  {journal} {\bibinfo  {journal} {Phys.
  Rev. B}\ }\textbf {\bibinfo {volume} {84}},\ \bibinfo {pages} {241105(R)}
  (\bibinfo {year} {2011})}\BibitemShut {NoStop}%
\bibitem [{\citenamefont {Rachel}\ and\ \citenamefont
  {Le~Hur}(2010)}]{Rachel10}%
  \BibitemOpen
  \bibfield  {author} {\bibinfo {author} {\bibfnamefont {S.}~\bibnamefont
  {Rachel}}\ and\ \bibinfo {author} {\bibfnamefont {K.}~\bibnamefont
  {Le~Hur}},\ }\href {\doibase 10.1103/PhysRevB.82.075106} {\bibfield
  {journal} {\bibinfo  {journal} {Phys. Rev. B}\ }\textbf {\bibinfo {volume}
  {82}},\ \bibinfo {pages} {075106} (\bibinfo {year} {2010})}\BibitemShut
  {NoStop}%
\bibitem [{\citenamefont {Yamaji}\ and\ \citenamefont
  {Imada}(2011)}]{Yamaji11}%
  \BibitemOpen
  \bibfield  {author} {\bibinfo {author} {\bibfnamefont {Y.}~\bibnamefont
  {Yamaji}}\ and\ \bibinfo {author} {\bibfnamefont {M.}~\bibnamefont {Imada}},\
  }\href {\doibase 10.1103/PhysRevB.83.205122} {\bibfield  {journal} {\bibinfo
  {journal} {Phys. Rev. B}\ }\textbf {\bibinfo {volume} {83}},\ \bibinfo
  {pages} {205122} (\bibinfo {year} {2011})}\BibitemShut {NoStop}%
\bibitem [{\citenamefont {Zheng}\ \emph {et~al.}(2011)\citenamefont {Zheng},
  \citenamefont {Zhang},\ and\ \citenamefont {Wu}}]{Zheng11}%
  \BibitemOpen
  \bibfield  {author} {\bibinfo {author} {\bibfnamefont {D.}~\bibnamefont
  {Zheng}}, \bibinfo {author} {\bibfnamefont {G.-M.}\ \bibnamefont {Zhang}}, \
  and\ \bibinfo {author} {\bibfnamefont {C.}~\bibnamefont {Wu}},\ }\href
  {\doibase 10.1103/PhysRevB.84.205121} {\bibfield  {journal} {\bibinfo
  {journal} {Phys. Rev. B}\ }\textbf {\bibinfo {volume} {84}},\ \bibinfo
  {pages} {205121} (\bibinfo {year} {2011})}\BibitemShut {NoStop}%
\bibitem [{\citenamefont {Yu}\ \emph {et~al.}(2011)\citenamefont {Yu},
  \citenamefont {Xie},\ and\ \citenamefont {Li}}]{Yu11}%
  \BibitemOpen
  \bibfield  {author} {\bibinfo {author} {\bibfnamefont {S.-L.}\ \bibnamefont
  {Yu}}, \bibinfo {author} {\bibfnamefont {X.~C.}\ \bibnamefont {Xie}}, \ and\
  \bibinfo {author} {\bibfnamefont {J.-X.}\ \bibnamefont {Li}},\ }\href
  {\doibase 10.1103/PhysRevLett.107.010401} {\bibfield  {journal} {\bibinfo
  {journal} {Phys. Rev. Lett.}\ }\textbf {\bibinfo {volume} {107}},\ \bibinfo
  {pages} {010401} (\bibinfo {year} {2011})}\BibitemShut {NoStop}%
\bibitem [{\citenamefont {Griset}\ and\ \citenamefont {Xu}(2012)}]{Griset12}%
  \BibitemOpen
  \bibfield  {author} {\bibinfo {author} {\bibfnamefont {C.}~\bibnamefont
  {Griset}}\ and\ \bibinfo {author} {\bibfnamefont {C.}~\bibnamefont {Xu}},\
  }\href {\doibase 10.1103/PhysRevB.85.045123} {\bibfield  {journal} {\bibinfo
  {journal} {Phys. Rev. B}\ }\textbf {\bibinfo {volume} {85}},\ \bibinfo
  {pages} {045123} (\bibinfo {year} {2012})}\BibitemShut {NoStop}%
\bibitem [{\citenamefont {Hohenadler}\ \emph {et~al.}(2011)\citenamefont
  {Hohenadler}, \citenamefont {Lang},\ and\ \citenamefont
  {Assaad}}]{Hohenadler11}%
  \BibitemOpen
  \bibfield  {author} {\bibinfo {author} {\bibfnamefont {M.}~\bibnamefont
  {Hohenadler}}, \bibinfo {author} {\bibfnamefont {T.~C.}\ \bibnamefont
  {Lang}}, \ and\ \bibinfo {author} {\bibfnamefont {F.~F.}\ \bibnamefont
  {Assaad}},\ }\href {\doibase 10.1103/PhysRevLett.106.100403} {\bibfield
  {journal} {\bibinfo  {journal} {Phys. Rev. Lett.}\ }\textbf {\bibinfo
  {volume} {106}},\ \bibinfo {pages} {100403} (\bibinfo {year}
  {2011})}\BibitemShut {NoStop}%
\bibitem [{\citenamefont {Hohenadler}\ \emph {et~al.}(2012)\citenamefont
  {Hohenadler}, \citenamefont {Meng}, \citenamefont {Lang}, \citenamefont
  {Wessel}, \citenamefont {Muramatsu},\ and\ \citenamefont
  {Assaad}}]{Hohenadler12}%
  \BibitemOpen
  \bibfield  {author} {\bibinfo {author} {\bibfnamefont {M.}~\bibnamefont
  {Hohenadler}}, \bibinfo {author} {\bibfnamefont {Z.~Y.}\ \bibnamefont
  {Meng}}, \bibinfo {author} {\bibfnamefont {T.~C.}\ \bibnamefont {Lang}},
  \bibinfo {author} {\bibfnamefont {S.}~\bibnamefont {Wessel}}, \bibinfo
  {author} {\bibfnamefont {A.}~\bibnamefont {Muramatsu}}, \ and\ \bibinfo
  {author} {\bibfnamefont {F.~F.}\ \bibnamefont {Assaad}},\ }\href {\doibase
  10.1103/PhysRevB.85.115132} {\bibfield  {journal} {\bibinfo  {journal} {Phys.
  Rev. B}\ }\textbf {\bibinfo {volume} {85}},\ \bibinfo {pages} {115132}
  (\bibinfo {year} {2012})}\BibitemShut {NoStop}%
\bibitem [{\citenamefont {Reuther}\ \emph {et~al.}(2012)\citenamefont
  {Reuther}, \citenamefont {Thomale},\ and\ \citenamefont
  {Rachel}}]{Reuther12}%
  \BibitemOpen
  \bibfield  {author} {\bibinfo {author} {\bibfnamefont {J.}~\bibnamefont
  {Reuther}}, \bibinfo {author} {\bibfnamefont {R.}~\bibnamefont {Thomale}}, \
  and\ \bibinfo {author} {\bibfnamefont {S.}~\bibnamefont {Rachel}},\ }\href
  {\doibase 10.1103/PhysRevB.86.155127} {\bibfield  {journal} {\bibinfo
  {journal} {Phys. Rev. B}\ }\textbf {\bibinfo {volume} {86}},\ \bibinfo
  {pages} {155127} (\bibinfo {year} {2012})}\BibitemShut {NoStop}%
\bibitem [{\citenamefont {Laubach}\ \emph {et~al.}(2014)\citenamefont
  {Laubach}, \citenamefont {Reuther}, \citenamefont {Thomale},\ and\
  \citenamefont {Rachel}}]{Laubach14}%
  \BibitemOpen
  \bibfield  {author} {\bibinfo {author} {\bibfnamefont {M.}~\bibnamefont
  {Laubach}}, \bibinfo {author} {\bibfnamefont {J.}~\bibnamefont {Reuther}},
  \bibinfo {author} {\bibfnamefont {R.}~\bibnamefont {Thomale}}, \ and\
  \bibinfo {author} {\bibfnamefont {S.}~\bibnamefont {Rachel}},\ }\href
  {\doibase 10.1103/PhysRevB.90.165136} {\bibfield  {journal} {\bibinfo
  {journal} {Phys. Rev. B}\ }\textbf {\bibinfo {volume} {90}},\ \bibinfo
  {pages} {165136} (\bibinfo {year} {2014})}\BibitemShut {NoStop}%
\bibitem [{\citenamefont {Shao}\ \emph {et~al.}(2021)\citenamefont {Shao},
  \citenamefont {Castro}, \citenamefont {Hu},\ and\ \citenamefont
  {Mondaini}}]{Shao21}%
  \BibitemOpen
  \bibfield  {author} {\bibinfo {author} {\bibfnamefont {C.}~\bibnamefont
  {Shao}}, \bibinfo {author} {\bibfnamefont {E.~V.}\ \bibnamefont {Castro}},
  \bibinfo {author} {\bibfnamefont {S.}~\bibnamefont {Hu}}, \ and\ \bibinfo
  {author} {\bibfnamefont {R.}~\bibnamefont {Mondaini}},\ }\href {\doibase
  10.1103/PhysRevB.103.035125} {\bibfield  {journal} {\bibinfo  {journal}
  {Phys. Rev. B}\ }\textbf {\bibinfo {volume} {103}},\ \bibinfo {pages}
  {035125} (\bibinfo {year} {2021})}\BibitemShut {NoStop}%
\bibitem [{\citenamefont {Raghu}\ \emph {et~al.}(2008)\citenamefont {Raghu},
  \citenamefont {Qi}, \citenamefont {Honerkamp},\ and\ \citenamefont
  {Zhang}}]{Raghu08}%
  \BibitemOpen
  \bibfield  {author} {\bibinfo {author} {\bibfnamefont {S.}~\bibnamefont
  {Raghu}}, \bibinfo {author} {\bibfnamefont {X.-L.}\ \bibnamefont {Qi}},
  \bibinfo {author} {\bibfnamefont {C.}~\bibnamefont {Honerkamp}}, \ and\
  \bibinfo {author} {\bibfnamefont {S.-C.}\ \bibnamefont {Zhang}},\ }\href
  {\doibase 10.1103/PhysRevLett.100.156401} {\bibfield  {journal} {\bibinfo
  {journal} {Phys. Rev. Lett.}\ }\textbf {\bibinfo {volume} {100}},\ \bibinfo
  {pages} {156401} (\bibinfo {year} {2008})}\BibitemShut {NoStop}%
\bibitem [{\citenamefont {Wen}\ \emph {et~al.}(2010)\citenamefont {Wen},
  \citenamefont {R\"uegg}, \citenamefont {Wang},\ and\ \citenamefont
  {Fiete}}]{Wen10}%
  \BibitemOpen
  \bibfield  {author} {\bibinfo {author} {\bibfnamefont {J.}~\bibnamefont
  {Wen}}, \bibinfo {author} {\bibfnamefont {A.}~\bibnamefont {R\"uegg}},
  \bibinfo {author} {\bibfnamefont {C.-C.~J.}\ \bibnamefont {Wang}}, \ and\
  \bibinfo {author} {\bibfnamefont {G.~A.}\ \bibnamefont {Fiete}},\ }\href
  {\doibase 10.1103/PhysRevB.82.075125} {\bibfield  {journal} {\bibinfo
  {journal} {Phys. Rev. B}\ }\textbf {\bibinfo {volume} {82}},\ \bibinfo
  {pages} {075125} (\bibinfo {year} {2010})}\BibitemShut {NoStop}%
\bibitem [{\citenamefont {Weeks}\ and\ \citenamefont {Franz}(2010)}]{Weeks10}%
  \BibitemOpen
  \bibfield  {author} {\bibinfo {author} {\bibfnamefont {C.}~\bibnamefont
  {Weeks}}\ and\ \bibinfo {author} {\bibfnamefont {M.}~\bibnamefont {Franz}},\
  }\href {\doibase 10.1103/PhysRevB.81.085105} {\bibfield  {journal} {\bibinfo
  {journal} {Phys. Rev. B}\ }\textbf {\bibinfo {volume} {81}},\ \bibinfo
  {pages} {085105} (\bibinfo {year} {2010})}\BibitemShut {NoStop}%
\bibitem [{\citenamefont {R\"uegg}\ and\ \citenamefont
  {Fiete}(2011)}]{Ruegg11}%
  \BibitemOpen
  \bibfield  {author} {\bibinfo {author} {\bibfnamefont {A.}~\bibnamefont
  {R\"uegg}}\ and\ \bibinfo {author} {\bibfnamefont {G.~A.}\ \bibnamefont
  {Fiete}},\ }\href {\doibase 10.1103/PhysRevB.84.201103} {\bibfield  {journal}
  {\bibinfo  {journal} {Phys. Rev. B}\ }\textbf {\bibinfo {volume} {84}},\
  \bibinfo {pages} {201103(R)} (\bibinfo {year} {2011})}\BibitemShut {NoStop}%
\bibitem [{\citenamefont {Yang}\ \emph {et~al.}(2011)\citenamefont {Yang},
  \citenamefont {Zhu}, \citenamefont {Xiao}, \citenamefont {Okamoto},
  \citenamefont {Wang},\ and\ \citenamefont {Ran}}]{Yang11}%
  \BibitemOpen
  \bibfield  {author} {\bibinfo {author} {\bibfnamefont {K.-Y.}\ \bibnamefont
  {Yang}}, \bibinfo {author} {\bibfnamefont {W.}~\bibnamefont {Zhu}}, \bibinfo
  {author} {\bibfnamefont {D.}~\bibnamefont {Xiao}}, \bibinfo {author}
  {\bibfnamefont {S.}~\bibnamefont {Okamoto}}, \bibinfo {author} {\bibfnamefont
  {Z.}~\bibnamefont {Wang}}, \ and\ \bibinfo {author} {\bibfnamefont
  {Y.}~\bibnamefont {Ran}},\ }\href {\doibase 10.1103/PhysRevB.84.201104}
  {\bibfield  {journal} {\bibinfo  {journal} {Phys. Rev. B}\ }\textbf {\bibinfo
  {volume} {84}},\ \bibinfo {pages} {201104(R)} (\bibinfo {year}
  {2011})}\BibitemShut {NoStop}%
\bibitem [{\citenamefont {Budich}\ \emph {et~al.}(2012)\citenamefont {Budich},
  \citenamefont {Thomale}, \citenamefont {Li}, \citenamefont {Laubach},\ and\
  \citenamefont {Zhang}}]{Budich12}%
  \BibitemOpen
  \bibfield  {author} {\bibinfo {author} {\bibfnamefont {J.~C.}\ \bibnamefont
  {Budich}}, \bibinfo {author} {\bibfnamefont {R.}~\bibnamefont {Thomale}},
  \bibinfo {author} {\bibfnamefont {G.}~\bibnamefont {Li}}, \bibinfo {author}
  {\bibfnamefont {M.}~\bibnamefont {Laubach}}, \ and\ \bibinfo {author}
  {\bibfnamefont {S.-C.}\ \bibnamefont {Zhang}},\ }\href {\doibase
  10.1103/PhysRevB.86.201407} {\bibfield  {journal} {\bibinfo  {journal} {Phys.
  Rev. B}\ }\textbf {\bibinfo {volume} {86}},\ \bibinfo {pages} {201407(R)}
  (\bibinfo {year} {2012})}\BibitemShut {NoStop}%
\bibitem [{\citenamefont {Dauphin}\ \emph {et~al.}(2012)\citenamefont
  {Dauphin}, \citenamefont {M\"uller},\ and\ \citenamefont
  {Martin-Delgado}}]{Dauphin12}%
  \BibitemOpen
  \bibfield  {author} {\bibinfo {author} {\bibfnamefont {A.}~\bibnamefont
  {Dauphin}}, \bibinfo {author} {\bibfnamefont {M.}~\bibnamefont {M\"uller}}, \
  and\ \bibinfo {author} {\bibfnamefont {M.~A.}\ \bibnamefont
  {Martin-Delgado}},\ }\href {\doibase 10.1103/PhysRevA.86.053618} {\bibfield
  {journal} {\bibinfo  {journal} {Phys. Rev. A}\ }\textbf {\bibinfo {volume}
  {86}},\ \bibinfo {pages} {053618} (\bibinfo {year} {2012})}\BibitemShut
  {NoStop}%
\bibitem [{\citenamefont {Wang}\ \emph {et~al.}(2012)\citenamefont {Wang},
  \citenamefont {Dai},\ and\ \citenamefont {Xie}}]{LeiWang12}%
  \BibitemOpen
  \bibfield  {author} {\bibinfo {author} {\bibfnamefont {L.}~\bibnamefont
  {Wang}}, \bibinfo {author} {\bibfnamefont {X.}~\bibnamefont {Dai}}, \ and\
  \bibinfo {author} {\bibfnamefont {X.~C.}\ \bibnamefont {Xie}},\ }\href
  {http://stacks.iop.org/0295-5075/98/i=5/a=57001} {\bibfield  {journal}
  {\bibinfo  {journal} {Europhys. Lett.}\ }\textbf {\bibinfo {volume} {98}},\
  \bibinfo {pages} {57001} (\bibinfo {year} {2012})}\BibitemShut {NoStop}%
\bibitem [{\citenamefont {Yoshida}\ \emph {et~al.}(2014)\citenamefont
  {Yoshida}, \citenamefont {Peters}, \citenamefont {Fujimoto},\ and\
  \citenamefont {Kawakami}}]{Yoshida14}%
  \BibitemOpen
  \bibfield  {author} {\bibinfo {author} {\bibfnamefont {T.}~\bibnamefont
  {Yoshida}}, \bibinfo {author} {\bibfnamefont {R.}~\bibnamefont {Peters}},
  \bibinfo {author} {\bibfnamefont {S.}~\bibnamefont {Fujimoto}}, \ and\
  \bibinfo {author} {\bibfnamefont {N.}~\bibnamefont {Kawakami}},\ }\href
  {\doibase 10.1103/PhysRevLett.112.196404} {\bibfield  {journal} {\bibinfo
  {journal} {Phys. Rev. Lett.}\ }\textbf {\bibinfo {volume} {112}},\ \bibinfo
  {pages} {196404} (\bibinfo {year} {2014})}\BibitemShut {NoStop}%
\bibitem [{\citenamefont {Oka}\ and\ \citenamefont {Aoki}(2009)}]{Oka09}%
  \BibitemOpen
  \bibfield  {author} {\bibinfo {author} {\bibfnamefont {T.}~\bibnamefont
  {Oka}}\ and\ \bibinfo {author} {\bibfnamefont {H.}~\bibnamefont {Aoki}},\
  }\href {\doibase 10.1103/PhysRevB.79.081406} {\bibfield  {journal} {\bibinfo
  {journal} {Phys. Rev. B}\ }\textbf {\bibinfo {volume} {79}},\ \bibinfo
  {pages} {081406(R)} (\bibinfo {year} {2009})}\BibitemShut {NoStop}%
\bibitem [{\citenamefont {Inoue}\ and\ \citenamefont {Tanaka}(2010)}]{Inoue10}%
  \BibitemOpen
  \bibfield  {author} {\bibinfo {author} {\bibfnamefont {J.-i.}\ \bibnamefont
  {Inoue}}\ and\ \bibinfo {author} {\bibfnamefont {A.}~\bibnamefont {Tanaka}},\
  }\href {\doibase 10.1103/PhysRevLett.105.017401} {\bibfield  {journal}
  {\bibinfo  {journal} {Phys. Rev. Lett.}\ }\textbf {\bibinfo {volume} {105}},\
  \bibinfo {pages} {017401} (\bibinfo {year} {2010})}\BibitemShut {NoStop}%
\bibitem [{\citenamefont {Kitagawa}\ \emph {et~al.}(2011)\citenamefont
  {Kitagawa}, \citenamefont {Oka}, \citenamefont {Brataas}, \citenamefont
  {Fu},\ and\ \citenamefont {Demler}}]{Kitagawa11}%
  \BibitemOpen
  \bibfield  {author} {\bibinfo {author} {\bibfnamefont {T.}~\bibnamefont
  {Kitagawa}}, \bibinfo {author} {\bibfnamefont {T.}~\bibnamefont {Oka}},
  \bibinfo {author} {\bibfnamefont {A.}~\bibnamefont {Brataas}}, \bibinfo
  {author} {\bibfnamefont {L.}~\bibnamefont {Fu}}, \ and\ \bibinfo {author}
  {\bibfnamefont {E.}~\bibnamefont {Demler}},\ }\href {\doibase
  10.1103/PhysRevB.84.235108} {\bibfield  {journal} {\bibinfo  {journal} {Phys.
  Rev. B}\ }\textbf {\bibinfo {volume} {84}},\ \bibinfo {pages} {235108}
  (\bibinfo {year} {2011})}\BibitemShut {NoStop}%
\bibitem [{\citenamefont {Lindner}\ \emph {et~al.}(2011)\citenamefont
  {Lindner}, \citenamefont {Refael},\ and\ \citenamefont
  {Galitski}}]{Lindner11}%
  \BibitemOpen
  \bibfield  {author} {\bibinfo {author} {\bibfnamefont {N.~H.}\ \bibnamefont
  {Lindner}}, \bibinfo {author} {\bibfnamefont {G.}~\bibnamefont {Refael}}, \
  and\ \bibinfo {author} {\bibfnamefont {V.}~\bibnamefont {Galitski}},\ }\href
  {\doibase 10.1038/nphys1926} {\bibfield  {journal} {\bibinfo  {journal}
  {Nature Physics}\ }\textbf {\bibinfo {volume} {7}},\ \bibinfo {pages} {490}
  (\bibinfo {year} {2011})}\BibitemShut {NoStop}%
\bibitem [{\citenamefont {Perez-Piskunow}\ \emph {et~al.}(2014)\citenamefont
  {Perez-Piskunow}, \citenamefont {Usaj}, \citenamefont {Balseiro},\ and\
  \citenamefont {Foa~Torres}}]{Perez-Piskunow14}%
  \BibitemOpen
  \bibfield  {author} {\bibinfo {author} {\bibfnamefont {P.~M.}\ \bibnamefont
  {Perez-Piskunow}}, \bibinfo {author} {\bibfnamefont {G.}~\bibnamefont
  {Usaj}}, \bibinfo {author} {\bibfnamefont {C.~A.}\ \bibnamefont {Balseiro}},
  \ and\ \bibinfo {author} {\bibfnamefont {L.~E.~F.}\ \bibnamefont
  {Foa~Torres}},\ }\href {\doibase 10.1103/PhysRevB.89.121401} {\bibfield
  {journal} {\bibinfo  {journal} {Phys. Rev. B}\ }\textbf {\bibinfo {volume}
  {89}},\ \bibinfo {pages} {121401(R)} (\bibinfo {year} {2014})}\BibitemShut
  {NoStop}%
\bibitem [{\citenamefont {Usaj}\ \emph {et~al.}(2014)\citenamefont {Usaj},
  \citenamefont {Perez-Piskunow}, \citenamefont {Foa~Torres},\ and\
  \citenamefont {Balseiro}}]{Usaj14}%
  \BibitemOpen
  \bibfield  {author} {\bibinfo {author} {\bibfnamefont {G.}~\bibnamefont
  {Usaj}}, \bibinfo {author} {\bibfnamefont {P.~M.}\ \bibnamefont
  {Perez-Piskunow}}, \bibinfo {author} {\bibfnamefont {L.~E.~F.}\ \bibnamefont
  {Foa~Torres}}, \ and\ \bibinfo {author} {\bibfnamefont {C.~A.}\ \bibnamefont
  {Balseiro}},\ }\href {\doibase 10.1103/PhysRevB.90.115423} {\bibfield
  {journal} {\bibinfo  {journal} {Phys. Rev. B}\ }\textbf {\bibinfo {volume}
  {90}},\ \bibinfo {pages} {115423} (\bibinfo {year} {2014})}\BibitemShut
  {NoStop}%
\bibitem [{\citenamefont {McIver}\ \emph {et~al.}(2020)\citenamefont {McIver},
  \citenamefont {Schulte}, \citenamefont {Stein}, \citenamefont {Matsuyama},
  \citenamefont {Jotzu}, \citenamefont {Meier},\ and\ \citenamefont
  {Cavalleri}}]{McIver20}%
  \BibitemOpen
  \bibfield  {author} {\bibinfo {author} {\bibfnamefont {J.~W.}\ \bibnamefont
  {McIver}}, \bibinfo {author} {\bibfnamefont {B.}~\bibnamefont {Schulte}},
  \bibinfo {author} {\bibfnamefont {F.-U.}\ \bibnamefont {Stein}}, \bibinfo
  {author} {\bibfnamefont {T.}~\bibnamefont {Matsuyama}}, \bibinfo {author}
  {\bibfnamefont {G.}~\bibnamefont {Jotzu}}, \bibinfo {author} {\bibfnamefont
  {G.}~\bibnamefont {Meier}}, \ and\ \bibinfo {author} {\bibfnamefont
  {A.}~\bibnamefont {Cavalleri}},\ }\href {\doibase 10.1038/s41567-019-0698-y}
  {\bibfield  {journal} {\bibinfo  {journal} {Nature Physics}\ }\textbf
  {\bibinfo {volume} {16}},\ \bibinfo {pages} {38} (\bibinfo {year}
  {2020})}\BibitemShut {NoStop}%
\bibitem [{\citenamefont {Shao}\ \emph {et~al.}(2019)\citenamefont {Shao},
  \citenamefont {Lu}, \citenamefont {Luo},\ and\ \citenamefont
  {Mondaini}}]{Shao19}%
  \BibitemOpen
  \bibfield  {author} {\bibinfo {author} {\bibfnamefont {C.}~\bibnamefont
  {Shao}}, \bibinfo {author} {\bibfnamefont {H.}~\bibnamefont {Lu}}, \bibinfo
  {author} {\bibfnamefont {H.-G.}\ \bibnamefont {Luo}}, \ and\ \bibinfo
  {author} {\bibfnamefont {R.}~\bibnamefont {Mondaini}},\ }\href {\doibase
  10.1103/PhysRevB.100.041114} {\bibfield  {journal} {\bibinfo  {journal}
  {Phys. Rev. B}\ }\textbf {\bibinfo {volume} {100}},\ \bibinfo {pages}
  {041114(R)} (\bibinfo {year} {2019})}\BibitemShut {NoStop}%
\bibitem [{\citenamefont {Niu}\ \emph {et~al.}(1985)\citenamefont {Niu},
  \citenamefont {Thouless},\ and\ \citenamefont {Wu}}]{Niu85}%
  \BibitemOpen
  \bibfield  {author} {\bibinfo {author} {\bibfnamefont {Q.}~\bibnamefont
  {Niu}}, \bibinfo {author} {\bibfnamefont {D.~J.}\ \bibnamefont {Thouless}}, \
  and\ \bibinfo {author} {\bibfnamefont {Y.-S.}\ \bibnamefont {Wu}},\ }\href
  {\doibase 10.1103/PhysRevB.31.3372} {\bibfield  {journal} {\bibinfo
  {journal} {Phys. Rev. B}\ }\textbf {\bibinfo {volume} {31}},\ \bibinfo
  {pages} {3372} (\bibinfo {year} {1985})}\BibitemShut {NoStop}%
\bibitem [{\citenamefont {Poilblanc}(1991)}]{Didier91}%
  \BibitemOpen
  \bibfield  {author} {\bibinfo {author} {\bibfnamefont {D.}~\bibnamefont
  {Poilblanc}},\ }\href {\doibase 10.1103/PhysRevB.44.9562} {\bibfield
  {journal} {\bibinfo  {journal} {Phys. Rev. B}\ }\textbf {\bibinfo {volume}
  {44}},\ \bibinfo {pages} {9562} (\bibinfo {year} {1991})}\BibitemShut
  {NoStop}%
\bibitem [{\citenamefont {Fukui}\ \emph {et~al.}(2005)\citenamefont {Fukui},
  \citenamefont {Hatsugai},\ and\ \citenamefont {Suzuki}}]{Fukui05}%
  \BibitemOpen
  \bibfield  {author} {\bibinfo {author} {\bibfnamefont {T.}~\bibnamefont
  {Fukui}}, \bibinfo {author} {\bibfnamefont {Y.}~\bibnamefont {Hatsugai}}, \
  and\ \bibinfo {author} {\bibfnamefont {H.}~\bibnamefont {Suzuki}},\ }\href
  {\doibase 10.1143/JPSJ.74.1674} {\bibfield  {journal} {\bibinfo  {journal}
  {J. Phys. Soc. Japan}\ }\textbf {\bibinfo {volume} {74}},\ \bibinfo {pages}
  {1674} (\bibinfo {year} {2005})}\BibitemShut {NoStop}%
\bibitem [{\citenamefont {Prelov$\check{\text{s}}$ek}\ and\ \citenamefont
  {Bon$\check{\text{c}}$a}()}]{Prelovsek}%
  \BibitemOpen
  \bibfield  {author} {\bibinfo {author} {\bibfnamefont {P.}~\bibnamefont
  {Prelov$\check{\text{s}}$ek}}\ and\ \bibinfo {author} {\bibfnamefont
  {J.}~\bibnamefont {Bon$\check{\text{c}}$a}},\ }\href@noop {} {\bibinfo
  {journal} {in {\it Strongly Correlated Systems-Numerical Methods}, edited by
  A. Avella and F. Mancini, Springer Series in Solid-State Sciences, Vol. 176
  (Springer, Berlin, 2013), pp. 1-30}\ }\BibitemShut {NoStop}%
\bibitem [{\citenamefont {Lu}\ \emph {et~al.}(2013)\citenamefont {Lu},
  \citenamefont {Bon{\v{c}}a},\ and\ \citenamefont {Tohyama}}]{Lu_2013}%
  \BibitemOpen
\bibfield  {journal} {  }\bibfield  {author} {\bibinfo {author} {\bibfnamefont
  {H.}~\bibnamefont {Lu}}, \bibinfo {author} {\bibfnamefont {J.}~\bibnamefont
  {Bon{\v{c}}a}}, \ and\ \bibinfo {author} {\bibfnamefont {T.}~\bibnamefont
  {Tohyama}},\ }\href {\doibase 10.1209/0295-5075/103/57005} {\bibfield
  {journal} {\bibinfo  {journal} {{EPL} (Europhysics Letters)}\ }\textbf
  {\bibinfo {volume} {103}},\ \bibinfo {pages} {57005} (\bibinfo {year}
  {2013})}\BibitemShut {NoStop}%
\end{thebibliography}
%

\end{document}